\documentclass[a4paper,12pt]{article}
\usepackage{jheppub}
\usepackage{amssymb}
\usepackage{amsmath}
\usepackage{graphicx}
\usepackage{multirow}
\usepackage{subfigure}
\usepackage{float}
\usepackage{graphics}
\usepackage{slashed}
\usepackage{booktabs}
\usepackage{color}
\usepackage[normalem]{ulem}
\def\be{\begin{equation}}
\def\ee{\end{equation}}
\newcommand{\bea}{\begin{equation} \begin{array}{c}}
\newcommand{\eea}{ \end{array} \end{equation}}

\def\as{\alpha_s}

\def\e{\epsilon}
\def\nno{\nonumber}

\def\ff{f\hspace{-0.4em}f}
\def\II{I\hspace{-0.7em}I}
\def\ptv{p_T^{\rm veto}}
\def\lt{L_{\perp}}
\def\sumint{\sum\hspace{-1.3em}\int}
\title{Resummation Prediction on Higgs and Vector Boson Associated Production with a Jet Veto at the LHC}

\author[a]{Ding Yu Shao,}
\author[a,b]{Chong Sheng Li,}
\author[a]{Hai Tao Li}

\affiliation[a]{School of Physics and State Key Laboratory of
Nuclear Physics and Technology, Peking University,\\Beijing 100871, China}
\affiliation[b]{Center for High Energy Physics,
Peking University,\\Beijing 100871, China}

\emailAdd{csli@pku.edu.cn}

\abstract{We investigate the resummation effects for the SM Higgs and vector boson associated production at the LHC with a jet veto in soft-collinear effective theory using ``collinear anomalous" formalism. We calculate the jet vetoed invariant mass distribution and the cross section for this process at Next-to-Next-to-Leading-Logarithmic level, which are matched to the QCD Next-to-Leading Order results, and compare the differences of the resummation effects with different jet veto $\ptv$ and jet radius $R$. Our results show that both resummation enhancement effects and the scale uncertainties decrease with the increasing of jet veto $\ptv$ and jet radius $R$, respectively. When $\ptv=25$ GeV and $R=0.4~(0.5)$, the resummation effects reduce the scale uncertainties of the Next-to-Leading Order jet vetoed cross sections to about $7\%~(6\%)$, which lead to increased confidence on the theoretical predictions. Besides, after including resummation effects, the PDF uncertainties of jet vetoed cross section are about $7\%$.}

\begin{document}
\bibliographystyle{unsrt}
\maketitle
\flushbottom

\section{INTRODUCTION}\label{s1}
Recently, both the ATLAS~\cite{Aad:2012tfa} and CMS~\cite{Chatrchyan:2012ufa} collaborations at the CERN Large Hadron Collider~(LHC) have found a Standard Model (SM) Higgs boson particle with a mass around $125~{\rm GeV}$ mainly through gluon-gluon fusion channel. However, by means of modern jet substructure methods, the associated production of Higgs boson $H$ and vector boson $V~(V = Z, W^{\pm})$ is also an important process to study the Higgs boson at the LHC.

The efforts of obtaining accurate theoretical predictions for $HV$ associated production at the hadron colliders have been for a long time. The Next-to-Leading-Order (NLO) QCD and Electro-Weak (EW) corrections have been performed in Refs.~\cite{Han:1991ia,Baer:1992vx,Ohnemus:1992bd,Kniehl:1990iva,
Ciccolini:2003jy}. Besides, the QCD Next-to-Next-to-Leading-Order (NNLO) corrections of the total inclusive cross section for $HV$ associated production were calculated in Refs.~\cite{Hamberg:1990np,Harlander:2002wh,Brein:2003wg}. The corresponding numerical results have been implemented in numerical code ${\rm VH}@{\rm NNLO}$~\cite{Brein:2012ne}, which is now available on the website. Recently, in Ref.~\cite{Ferrera:2011bk} the NNLO QCD corrections of exclusive cross section for $HW^\pm$ associated  production were completed based on the transverse momentum substraction formalism\cite{Catani:2007vq}. And the effects of NLO QCD corrections to both $HW^{\pm}$ associated production and subsequent decay of $H\to b\bar{b}$ were investigated in Ref.~\cite{Banfi:2012jh}. However, the completely NNLO QCD corrections for both $HV$ associated production and subsequent decay of $H\to b\bar{b}$ are still absent so far.

%The efforts of obtaining accurate theoretical predictions for $HV$ associated production at the hadron colliders have been for a long time. As far back as 20 years ago, the QCD Next-to-Leading-Order (NLO) corrections for $HV$ associated production were performed~\cite{Han:1991ia,Baer:1992vx,Ohnemus:1992bd}. The predictions of total inclusive cross section for $HZ$ associated production induced by gluon-gluon fusion at the Leading Order (LO) were also investigated in Ref.~\cite{Kniehl:1990iva}. Besides, the NLO Electro-Weak (EW) corrections were presented in Ref.\cite{Ciccolini:2003jy}. Moreover, the QCD Next-to-Next-to-Leading-Order (NNLO) corrections of the total inclusive cross section for $HV$ associated production through quark anti-quark annihilation were calculated in Ref.~\cite{Hamberg:1990np,Harlander:2002wh,Brein:2003wg}. The corresponding numerical results have been implemented in numerical code ${\rm VH}@{\rm NNLO}$~\cite{Brein:2012ne}, which is now available on the website. Recently, in Ref.~\cite{Ferrera:2011bk} the NNLO QCD corrections of exclusive cross section for $HW^\pm$ associated  production were completed based on the transverse momentum substraction formalism\cite{Catani:2007vq}. And the effects of NLO QCD corrections to both $HW^{\pm}$ associated production and subsequent decay of $H\to b\bar{b}$ were investigated in Ref.~\cite{Banfi:2012jh}. However, the NNLO QCD corrections for both $HV$ associated production and subsequent decay of $H\to b\bar{b}$ are still absent so far.

The process for Higgs boson production involve a number of jets associated radiation at hadron colliders. The Standard Model (SM) backgrounds process produce the similar signature with additional energetic jets. For example, the $HW^{\pm}$ associated production with Higgs decaying to $b\bar{b}$ has large QCD backgrounds at hadron colliders. When leptonic decay modes of $W^{\pm}$ is considered, the semi-leptonic decays of $t\bar{t}$ can become a significant irreducible background. Due to the fact that the SM top quark pair production has more hard jets from decay of top quark than the $HW^{\pm}$ process, a jet veto can be used to suppress $t\bar{t}$ background~\cite{Dawson:2012gs}. Thus, a veto on the additional undesired jets $p_T^{\rm jet}<\ptv$ is needed to distinguish the signal and background process, and improve the significance of $HW^{\pm}$ production.

Due to the presence of the jet veto $\ptv$, a small energy scale $\ptv$ is introduced into the physical process, which is about $20 \sim 30 $GeV. Therefore there exist large logarithmic terms $\ln^n\ptv/Q$ in the perturbative calculations at the all order where $Q$ denotes the hard scale in the process, and these large logarithms need be resummed for improving the accuracy of the theoretical predictions. By means of parton showers, the Leading-Logarithmic~(LL) predictions on the cross section with a jet veto are available \cite{Anastasiou:2008ik,Anastasiou:2009bt}. Besides, the event shape variables of beam thrust, $N-$jettiness and $E_T=\sum|\vec{p}_T|$ are used to implement a jet veto on additional emissions~\cite{Stewart:2009yx,Stewart:2010tn,Berger:2010xi,Stewart:2010pd,Papaefstathiou:2010bw,Tackmann:2012bt}. In the last year the jet veto efficiency in Higgs boson and Drell-Yan production at the hadron collider at the NLL level has been investigated with the CAESAR approach~\cite{Banfi:2004yd} in Ref.~\cite{Banfi:2012yh}. After that the all order factorization formula for single Higgs boson production with a jet veto $\ptv$ have been firstly derived at the leading power of $\lambda=\ptv/m_H$ with the soft-collinear effective theory (SCET)\cite{Bauer:2000yr,Bauer:2001yt,Beneke:2002ph} based on ``collinear anomaly" formalism~\cite{Becher:2010tm}, and the large double logarithmic terms have been resummed to NNLL order in Ref.~\cite{Becher:2012qa}. Then in Ref.~\cite{Banfi:2012jm}, the results of Ref.~\cite{Banfi:2004yd} combining the Drell-Yan like boson transverse momentum resummations~\cite{Bozzi:2003jy,Bozzi:2005wk,Bozzi:2010xn,Becher:2010tm} are used to obtain NNLL resummed jet veto efficiencies for Higgs boson and Drell-Yan production at hadron colliders. Very recently, the ${\rm N^2LL'}$+NNLO predictions on the jet veto cross section for single Higgs boson production have been investigated in Ref.~\cite{Becher:2013xia,Stewart:2013faa}. In Ref.~\cite{Becher:2013xia} the anomaly coefficient $d_2^{\rm veto}(R)$ was firstly calculated using the SCET and the two loop low energy matrix elements are extracted numerically. The main theoretical approximation comes from the lack of the anomaly coefficient $d_3^{\rm veto}(R)$ and the four loop cusp anomalous dimension. And in Ref.~\cite{Stewart:2013faa} the ``rapidity renormalization group" formalism~\cite{Chiu:2011qc,Chiu:2012ir} are used, where the NNLO soft function and the NNLO beam function are partly derived. The remaining contributions are numerically extracted. The main approximation also comes from unknown higher-order anomalous dimensions.

In this paper we investigate the resummation effects in $HV$ associated production at the hadron collider with a jet veto using SCET based on the ``collinear anomaly" formalism. We firstly calculate the Higgs and vector boson invariant mass distribution and the total cross section with a jet veto at the NNLL level, which are matched to the QCD NLO results. Nevertheless, the jet veto efficiency for $HV$ associated production have be approximated studied in Ref.~\cite{Dawson:2012gs}, where the jet veto cross section is defined as
\begin{eqnarray}
\sigma(p_{T, \, HV})=\int_{0}^{p_{T,HV}} d p_{T, \, HV} \frac{d\sigma}{d p_{T, \, HV}}.
\end{eqnarray}
Here $p_{T, \, HV}$ is the transverse momentum of $HV$ and $d\sigma/d p_{T, \, HV}$ is NLL+NLO $HV$ transverse momentum distribution. However the logarithmic terms at small $p_{T,HV}$ are different from those induced by jet veto $\ptv$ at the NNLL level, so those studies in Ref.~\cite{Dawson:2012gs} only give a qualitative analysis.

\indent The arrangement of this paper is as follows. In Sec.~\ref{s2} we derive the factorization formula for $HV$ associated production with a jet veto at the hadron collider. In Sec.~\ref{s3} we calculate the hard and beam matching coefficients at the NLO, and present Renormalization Group~(RG) improved differential cross section analytically. In Sec.~\ref{s4} we discuss the numerical results of cross section and the invariant mass distribution with a jet veto. We conclude in Sec.~\ref{s5}.

\section{FACTORIZATION IN SCET}\label{s2}
In this section we describe the derivation of factorization for $HV$ associated production with a jet veto in SCET based on the ``collinear anomaly" formalism. In Ref.~\cite{Dawson:2012gs} the threshold resummation of the total cross section and invariant mass distribution for $HV$ associated production in SCET has been investigated. However, the resummation for $HV$ associated production with a jet veto discussed in this paper is genuinely different from threshold resummation.  We consider the process of stable Higgs and vector boson associated production,
\begin{eqnarray}
 N_1(P_1)+N_2(P_2) \rightarrow H(p_3)+V(p_4)+X'(p_X),
\end{eqnarray}
where $X'$ is the final hadronic state passing jet veto $\ptv$. In the Born approximation $HV$ associated production is mainly induced by quark anti-quark annihilation,
\begin{eqnarray}
 q(p_1) + \bar{q}(p_2) \rightarrow H(p_3)+V(p_4),
\end{eqnarray}
where $p_1 = x_1 P_1$ and $p_2 = x_2 P_2$. We define the kinematic invariants,
\begin{eqnarray}
s=(P_1 + P_2)^2,~~\hat{s}=(p_1+p_2)^2,~~M^2=(p_3+p_4)^2.
\end{eqnarray}
In the presence of a jet veto $\ptv$, the kinematic region we are interested in is
\begin{eqnarray}
 \hat{s},M^2,m_H^2,m_V^2 \gg (\ptv)^2 \gg \Lambda_{\rm QCD}^2.
\end{eqnarray}
It is convenient to introduce two light-like reference vectors $n=(1,0,0,1)$ and $\bar{n}=(1,0,0,-1)$ along the beam axis and any four vector can be decomposed as
\begin{eqnarray}
p^{\mu} = n \cdot p \frac{\bar{n}^{\mu}}{2} + \bar{n}\cdot p \frac{n^{\mu}}{2} + p_{\perp}^{\mu} \equiv p_{+}^{\mu} + p_{-}^{\mu} + p_{\perp}^{\mu}.
\end{eqnarray}
Hence momentum $p^{\mu}$ can be denoted by $p^{\mu}=(p^{+},p^{-},p_{\perp})$. Different momentum modes relevant to our discussions are collinear mode $p^{\mu}_n \sim M(\lambda^2,1,\lambda)$, anti-collinear mode $p^{\mu}_{\bar{n}} \sim M(1,\lambda^2,\lambda)$ and soft mode $p^{\mu}_s \sim M(\lambda,\lambda,\lambda)$. Here $\lambda=\ptv/M$ is treated as a small expansion parameter. In order to handle these momentum regions, SCET is a very useful framework, which is very suitable to deal with the scattering processes with multiple scales.

For the Drell-Yan like process the chiral current operator for initial quark and anti-quark can be written as
\begin{align}
J^{\mu} = g_L^V \bar{q}_i\gamma^{\mu}P_L q_j + g_R^V \bar{q}_i \gamma^{\mu} P_R q_j
\end{align}
where the $i,j$ subscripts represent the flavors of quark and the couplings $g_{L(R)}^V$ for $W$ and $Z$ boson are separately
\begin{itemize}
  \item $g_L^{W} = \frac{V_{ij}}{\sqrt{2}S_w}$, ~~ $g_R^{W} = 0 $,
  \item $g_L^{Z} = \frac{I_f^3 - S_w^2 Q_f}{S_w C_w }\delta_{ij}$, ~~ $g_R^{Z} = - \frac{S_w}{C_w} Q_f \delta_{ij} $,
\end{itemize}
where $V_{ij}$ is the CKM matrix, $I_f^3$ is the third component of isospin and $Q_f$ is the electric charge for quark. Here $S_w=\sin\theta_{w}$ and $C_w=\cos\theta_{W}$, where $\theta_{W}$ is Weinberg angle. At the leading power of $\lambda$, the chiral current operators are matched onto SCET operators as
\begin{eqnarray}
 J^{\mu} \rightarrow C_V(-q^2 - i\epsilon, \mu^2)\left( g_L^V \bar{\chi}_{\bar{n}} S_{\bar n}^{\dagger} \gamma^{\mu} P_L S_{n} \chi_{n} + g_R^V \bar{\chi}_{\bar{n}} S_{\bar n}^{\dagger} \gamma^{\mu} P_R S_{n} \chi_{n}  \right).
\end{eqnarray}
Here $C_V$ is the hard matching coefficient and $\bar{\chi}_{n(\bar{n})}$ are the gauge invariant combinations of (anti-)collinear quark fields and Wilson lines in SCET. The soft degrees of freedom are  contained in the soft Wilson lines $S_{n(\bar{n})}$.

In order to define the jets at the hadron collider, the sequential recombination jet algorithms are used~\cite{Salam:2009jx}. The longitudinal boost invariant distance measures $d_{ij}$ and $d_{i \, B}$ are defined by
\begin{eqnarray}
 &&d_{ij}={\rm min}(p_{T \, i}^n, p_{T \, j}^n)\Delta R_{ij}/R,~~~~ \Delta R_{ij} = \sqrt{(y_i - y_j)^2 + (\phi_i - \phi_j^2)}, \\
 &&d_{i \, B} = p_{T \, i}^n,
\end{eqnarray}
where $R$ is the jet radius parameter. Here $n = -1, 0~{\rm and}~1$ represent the inclusive anti-$k_T$~\cite{Catani:1993hr,Ellis:1993tq}, Cambridge-Aachen~\cite{Dokshitzer:1997in,Wobisch:1998wt} and $k_T$~\cite{Cacciari:2008gp} jet algorithms, respectively. As is shown in Ref.~\cite{Becher:2012qa}, the different momentum modes (collinear, anti-collinear and soft) can not be grouped into the same jet after performing jet algorithms as long as jet radius parameter satisfies
\begin{eqnarray}
 \lambda \ll R \ll \ln \lambda,
\end{eqnarray}
where $R \sim \mathcal{O}(1)$ is assumed. Therefore the jet veto can be applied in collinear, anti-collinear and soft region, respectively. After factorizing the contributions from hard, collinear, anti-collinear, and soft degrees of freedom in the SCET, we can obtain the factorized differential cross section for the rapidity $Y$ and the invariant mass $M$ of Higgs and vector boson at the leading power of $\lambda$
\begin{eqnarray}\label{fac_cs_1}
 \frac{d\sigma(p_T^{\rm veto})}{d M^2 d Y} &=& \frac{\sigma_0}{s} \mathcal{H}(M^2,\mu^2) \mathcal{B}_{q/N_1}^n(\zeta_1, \ptv, \mu) \mathcal{B}_{\bar{q}/N_2}^{\bar{n}}(\zeta_2, \ptv, \mu) \mathcal{S}(\ptv,\mu) \nno \\
 &&+ ~~ (\,q \leftrightarrow \bar{q}\,),
\end{eqnarray}
were $\zeta_{1,\,2}=(M/\sqrt{s})e^{\pm Y}$ and $\sigma_0$ is the LO total cross section, and it is defined as
\begin{eqnarray}
 \sigma_0 = \frac{G_F^2 S_w^4 m_W^4}{36\pi M^2} g_{VVH}^2(g_L^2+g_R^2) \Lambda^{1/2}(m_V^2,m_H^2,M^2)\frac{\Lambda(m_V^2,m_H^2,M^2) + 12m_V^2/M^2}{(1 - m_V^2/M^2)^2}, \nno \\
\end{eqnarray}
with
\begin{eqnarray}
 \Lambda(x,y,z) = (1 - x/z - y/z)^2 - 4xy/z^2 .
\end{eqnarray}
 Here $m_V$ is the mass of vector boson, $G_F$ is Fermi constant, $g_{VVH}$ is the coupling between Higgs and vector boson, $g_{WWH}=1/S_w$ and $g_{ZZH}=1/(S_w C_w)$. In the Eq.~(\ref{fac_cs_1}), the hard function $\mathcal{H}$ is the absolute value squared of the hard matching coefficient $\mathcal{H}(M^2,\mu^2)=\left| C_{V}(-M^2-i\epsilon,\mu^2) \right|^2$, and the collinear matrix elements $\mathcal{B}_{q/N}^n$ correspond to the PDFs, which are defined as~\cite{Becher:2013xia}
\begin{eqnarray}\label{collinear_matrix}
 \mathcal{B}_{q/N}^n(z,\ptv,\mu)&=&\int\frac{d t}{2\pi}e^{-i z t \bar{n}\cdot p} \underset{X_n, {\rm reg}}{\sumint} \mathcal{M}_{\rm veto}(\ptv, R, \{ \underline{p_n} \}) \nno\\
 &&\times \langle N(p) | \bar{\chi}_n(t\bar{n}) | X_n \rangle \langle X_n | \chi_n(0) | N(p) \rangle.
\end{eqnarray}
Here the summation over the collinear states $X_n$ is constrained by the jet veto, and the corresponding constraints are included in the function $\mathcal{M}_{\rm veto}$, which depends on the collinear momentums $\{\underline{p_n}\}$. Similarly, the soft function is defined in terms of the vacuum matrix element of the product for the soft Wilson lines constrained by the jet veto as~\cite{Becher:2013xia}
\begin{eqnarray}
\mathcal{S}(\ptv,\mu) = \frac{1}{N_c} \underset{X_s, {\rm reg}}{\sumint} \mathcal{M}_{\rm veto}(\ptv, R, \{ \underline{p_s} \}) \langle 0 | [S_n^\dagger S_{\bar{n}}](0) | X_s \rangle \langle X_s | [S_{\bar{n}}^\dagger S_{n}](0) | 0 \rangle.  \nno \\
\end{eqnarray}
The definitions of the (anti-)collinear and soft functions involve light-cone singularities which are not regularized by dimensional regularization. These divergences can be regularized in various ways\cite{Becher:2010tm,Chiu:2011qc,Becher:2011dz,Echevarria:2012qe}, and the product of the (anti-)collinear and soft functions are free from the light-cone singularities. However, anomalous dependence on the hard scale $M$ remains, which was called ``collinear anomaly"~\cite{Becher:2010tm}.

\section{HARD FUNCTION AND BEAM FUNCTION}\label{s3}
\subsection{Hard function}
The hard matching coefficient $C_V(-M^2,\mu_h^2)$~(here and below the negative arguments are understood with a $-i\e$ prescription) can be obtained by matching the two quark operators in the full theory onto the operator in SCET, where the infrared divergences are subtracted in the $\overline{\rm MS}$ scheme. The two loop results for the $C_V(-M^2,\mu_h^2)$ have been available in Ref.~\cite{Becher:2007ty}. Up to NLO, it can be written as
\begin{eqnarray}
C_{V}(-M^2,\mu_h^2) = 1 + \frac{C_F\alpha_s(\mu_h^2)}{4\pi}\left( -L_H^2 + 3L_H - 8 + \frac{\pi^2}{6} \right),
\end{eqnarray}
where $L_H=\ln(-M^2/\mu_h^2)$. The RG equation for $C_V(-M^2,\mu^2)$ is governed by the anomalous-dimension, the structure of which has been predicted up to four-loop level for the case of massless partons \cite{Ahrens:2012qz}. The $C_V(-M^2,\mu^2)$ satisfies the RG equation
\begin{eqnarray}
\frac{d}{d\ln\mu}C_{V}(-M^2,\mu^2) = \left[ \Gamma_{\rm cusp}^F(\as)\ln\frac{-M^2}{\mu^2} + \gamma^V(\as) \right] C_{V}(-M^2,\mu^2),
\end{eqnarray}
where $\Gamma_{\rm cusp}^F(\as)$ is the cusp anomalous dimension, while $\gamma^V(\as)$ controls the single-logarithmic evolution. After solving the RG equation, we have the hard matching coefficient
\begin{align}
C_V(-M^2,\mu_f^2) &= \nno\\
 &\hspace{-4em} \exp\left[ 2S(\mu_h^2,\mu_f^2) - a_{\Gamma}(\mu_h^2,\mu_f^2)\ln\frac{-M^2}{\mu_h^2} - a_{\gamma^V}(\mu_h^2,\mu_f^2) \right] C_V(-M^2,\mu_h^2),
\end{align}
where $S(\nu^2,\mu^2)$ and $a_\Gamma(\nu^2,\mu^2)$ are defined as
\begin{eqnarray}
 S(\nu^2,\mu^2)&=&-\int_{\as(\nu^2)}^{\as(\mu^2)}d\alpha \frac{\Gamma_{\rm cusp}^F(\alpha)}{\beta(\alpha)}
 \int_{\as(\nu^2)}^{\alpha}\frac{d\alpha'}{\beta(\alpha')}, \\
 a_\Gamma(\nu^2,\mu^2)&=&-\int_{\as(\nu^2)}^{\as(\mu^2)}d\alpha \frac{\Gamma_{\rm cusp}^F(\alpha)}{\beta(\alpha)}.
\end{eqnarray}
$a_{\gamma^V}$ has a similar expression. Finally, the hard function is given by
\begin{eqnarray}
 \mathcal{H}(M^2,\mu_f^2)=\left| C_{V}(-M^2,\mu_f^2) \right|^2.
\end{eqnarray}
Up to NNLL level, we need three loop cusp anomalous dimension and two loop normal anomalous dimension, and their explicit expressions are collected in the Appendices of Ref.~\cite{Becher:2007ty}.

\subsection{Beam function}

In Ref.~\cite{Stewart:2009yx} a first study on the factorization theorem with beam function is performed. At hadron colliders if there exists experimental restrictions, which introduce a new kinematic scale on the hadronic final states, then the factorization does not yield standard PDFs for the initial states. Thus beam function is necessary to properly describe the jets from initial states.

The collinear matrix element $\mathcal{B}_{q/N}^n(z,\ptv,\mu)$ defined in Eq.(\ref{collinear_matrix}) are intrinsically non-perturbative objects. In the limit $\ptv \gg \Lambda_{\rm QCD}$, they can be matched onto the standard Parton Distribution Functions~(PDFs) via~\cite{Becher:2012qa}
\begin{eqnarray}\label{beam_match}
 \mathcal{B}_{q/N}^n(\zeta,\ptv,\mu) = \sum_{i = g, q, \bar{q}}\int_{\zeta}^1\frac{d z}{z} \mathcal{I}_{q \leftarrow i}(z,\ptv,\mu) f_{i/N}(\zeta/z,\mu),
\end{eqnarray}
where the beam function $\mathcal{I}_{q \leftarrow i}(z,\ptv,\mu)$ can be calculated up to QCD NLO and we collect those results in Appendix~\ref{a1} for the convenience. The product of initial state beam functions can be factorized as
\begin{align}\label{beam_fac_1}
 \left[\mathcal{I}_{q \leftarrow i}(z_1,\ptv,\mu_f) \mathcal{I}_{\bar{q} \leftarrow j}(z_2,\ptv,\mu_f)\right]_{q^2=M^2} &=  \nno \\
 & \hspace{-7em} \left( \frac{M}{\ptv} \right)^{-2 F_{q\bar{q}}(\ptv, \, \mu_f)} I_{q \leftarrow i}(z_1,\ptv,\mu_f) I_{\bar{q} \leftarrow j}(z_2,\ptv,\mu_f),
\end{align}
where the anomalous dependence on $M$ is factorized out and is controlled by the function $F_{q\bar{q}}$, while the function $I_{q \leftarrow i}$ is independent on the hard scale $M$. The RG equation for $F_{q\bar{q}}$ can be written as
\begin{eqnarray}\label{Fqq_rg}
 &&\frac{d}{d\ln \mu}F_{q\bar{q}}(\ptv, \mu) = 2\Gamma_{\rm cusp}^F(\as).
\end{eqnarray}
After solving this RG equation, we can obtain $F_{q\bar{q}}$ up to two loop as
\begin{eqnarray}
F_{q\bar{q}}(p_T^{\rm veto},\mu_f) &=& a_s\left[\Gamma_0^F L_{\perp} + d_1^{\rm veto}(R)\right] + a_s^2\left[\Gamma_0^{\rm F} \beta_0 \frac{L_{\perp}^2}{2} + \Gamma_1^{\rm F} L_{\perp} + d_2^{\rm veto}(R) \right], \nno \\
\end{eqnarray}
where the anomaly coefficient $d_i^{\rm veto}(R)$ can be extracted from fixed order calculations of beam function. In order to cancel large logarithms dependence in function $I_{q \leftarrow i}$, the double logarithmic terms in the $I_{q \leftarrow i}$ functions are exponentiated via
\begin{align}\label{beam_fac_2}
  \overline{I}_{q \leftarrow i}(z,\ptv,\mu_f) = e^{-h_F(\ptv, \, \mu_f)} I_{q \leftarrow i}(z,\ptv,\mu_f),
\end{align}
where the RG equation for $h_F$ can be written as
\begin{eqnarray}
 &&\frac{d}{d\ln\mu} h_F(\ptv , \, \mu) = 2\Gamma_{\rm cusp}^F(\as)\ln\frac{\mu}{\ptv} - 2\gamma^q(\as).
\end{eqnarray}
Here $\gamma^q$ is the anomalous dimension of collinear quark field. The solution of this RG equation for $h_F$ is given by
\begin{eqnarray}
h_F(p_T^{\rm veto},\mu_f) &=& a_s\left( \Gamma_0^F\frac{L_{\perp}^2}{4} - \gamma_0^q L_{\perp} \right),
\end{eqnarray}
where the normalization condition of $h_F(\ptv,\ptv) \equiv 0$ is chosen. Now, the RG equation for the matching function $\overline{I}_{q \leftarrow i}(z,\ptv,\mu_f)$ can be written as
\begin{eqnarray}\label{I_rge}
 \frac{d}{d\ln\mu}\overline{I}_{q \leftarrow i}(z, \ptv, \mu) = - \sum_{j}\int_z^1\frac{d\zeta}{\zeta}\overline{I}_{q \leftarrow j}(\zeta, \ptv, \mu)\mathcal{P}_{j \leftarrow i}(z/\zeta,\as).
\end{eqnarray}
Here $\mathcal{P}_{j \leftarrow i}$ are the DGLAP splitting functions. Obviously, the new functions $\overline{I}_{q \leftarrow j}$ evolve exactly following the DGLAP equations with an opposite sign. Solving the RG equation~(\ref{I_rge}), up to the NLO, we have
\begin{eqnarray}
 \overline{I}_{q \leftarrow i}(z, \ptv, \mu_f) = \delta(1-z)\delta_{qi} + a_s\left[ -\mathcal{P}_{q\leftarrow i}^{(1)}(z)\frac{\lt}{2} + \mathcal{R}_{q\leftarrow i}(z) \right].
\end{eqnarray}
Here we define $a_s \equiv \as/(4\pi)$, $L_{\perp} \equiv 2 \ln(\mu_f/\ptv)$. After calculating complete one loop function $\mathcal{I}_{q \leftarrow i}(z,\ptv,\mu)\,$, we have
\begin{eqnarray}\label{beam_1}
 d_1^{\rm veto}(R) &=& 0, \\
 \mathcal{R}_{q \leftarrow q}(z) &=& C_F\left[2(1-z) - \frac{\pi^2}{6}\delta(1-z)\right], \\
 \mathcal{R}_{q \leftarrow g}(z) &=& 4 T_F z(1-z).
\end{eqnarray}
The two loop coefficient $d_2^{\rm veto}(R)$ expanded as small $R$ has been analytically calculated in Ref.~\cite{Becher:2013xia}, and it has the form
\begin{eqnarray}
d_2^{\rm veto} = d_2^{\,q} - 8\Gamma_0^{\rm F} f(R),
\end{eqnarray}
where $d_2^q$ is the corresponding coefficient in the small transverse momentum resummation for Drell-Yan process and is given by
\begin{eqnarray}
d_2^{\,q} = \Gamma_0^F \left[\left( \frac{202}{27} - 7 \zeta_3 \right)C_A -\frac{56}{27} T_F n_f \right],
\end{eqnarray}
and the function $f(R)$ can also be numerically extracted from Ref.~\cite{Banfi:2012jm,Banfi:2004yd}, which agrees well with the analytical expression in Ref.~\cite{Becher:2013xia}, which is
\begin{eqnarray}\label{fR}
f(R) &=& -(1.09626 C_A + 0.1768 n_f T_F) \ln R + (0.6072 C_A - 0.0308 T_F n_f) \nno \\
&& + ( 0.2639 C_A - 0.8225 C_F + 0.02207 T_F n_f )R^2  \nno \\
&&- (0.0226 C_A - 0.0625 C_F + 0.0004 T_F n_f)R^4 + \cdots .
\end{eqnarray}

\subsection{RG improved cross section}
Based on the regularization scheme in Ref.~\cite{Becher:2011dz}, the soft function $\mathcal{S}(\ptv,\mu)\equiv1$ to all order because the integrals of soft function are scaleless in the high order perturbative calculations. Therefore, after integrating the the rapidity variable $Y$, we finally have the resummed cross section
\begin{eqnarray}\label{fac_cs_3}
 \frac{d\sigma(p_T^{\rm veto})}{d M^2} &=& \frac{\sigma_0}{s} \overline{H}(M,\ptv) \int_{\tau}^1 \frac{d z}{z} \, \overline{\II}_{ij}(z, p_T^{\rm veto},\mu_f) \ff_{ij}\left( \frac{\tau}{z},\mu_f \right).
\end{eqnarray}
where we have defined the RG invariant hard function as
\begin{eqnarray}
 &&\overline{H}(M,\ptv) = \mathcal{H}(M^2,\mu_f^2) \left( \frac{M}{\ptv} \right)^{-2 F_{q\bar{q}}(\ptv, \, \mu_f)} e^{2h_F(\ptv, \, \mu_f)},
\end{eqnarray}
and the convolutions of $\overline{I}_{q \leftarrow i}$ and PDF are given by
\begin{align}
 &\overline{\II}_{ij}(z, p_T^{\rm veto},\mu_f) = \int_{z}^1\frac{d u}{u} \overline{I}_{q \leftarrow i}(u,p_T^{\rm veto},\mu_f) \overline{I}_{\bar{q} \leftarrow j}(z/u,p_T^{\rm veto},\mu_f) + (q \leftrightarrow \bar{q} ), \\
 &\hspace{-1.8em} {\rm and} \nno \\
 &\ff_{ij}\left(y,\mu_f\right) = \int_y^1 \frac{dx}{x}f_i(x,\mu_f)f_j\left(\frac{\tau}{x z},\mu_f\right).
\end{align}
respectively. Here $(ij)=(q\bar{q}), (qg)$ and $(g\bar{q})$. In order to give precise predictions, we resum the leading singular terms to all orders and include the nonsingular terms, which are suppressed by powers of $\lambda$, up to NLO. Finally, we obtain the RG improved differential cross section as
\begin{eqnarray}
 \frac{d\sigma^{\rm NLO+NNLL}(\ptv)}{d M^2} = \frac{d\sigma^{ \rm NNLL}(\ptv)}{d M^2} + \left[ \frac{d\sigma^{\rm NLO}}{d M^2} - \frac{d\sigma^{\rm NNLL}(\ptv)}{d M^2} \right]_{\rm expand~to~NLO} . \nno \\
\end{eqnarray}

In this paper our main goal is to derive the factorization expressions and perform the resummation calculations for $HV$ production with a jet veto. The numerical results of the differential NNLO QCD predictions for $HW^{\pm}$ are shown in Ref.~\cite{Ferrera:2011bk}, but their numerical code has not been published. Repeating the complete numerical NNLO QCD calculations is beyond the scope of the this paper. Therefore, we will only include the QCD NLO results in this paper.

The EW gauge boson pair $W^{+}W^{-}$ production with a jet veto at the LHC is a main SM background for the jet veto Higgs boson production channel $gg \rightarrow H \rightarrow W^{+}W^{-}$, and thus it is also significant to perform the resummation calculations for $W^{+}W^{-}$ production with a jet veto. Our results can be easily extended to $W^{+}W^{-}$ production with a jet veto, and the only differences come from LO cross section and the scale independent terms in the hard functions, which are collected in Refs.~\cite{Frixione:1993yp,Dawson:2013lya,Wang:2013qua}.

\section{NUMERICAL RESULTS}\label{s4}
In this section, we discuss the numerical results for the $HV$ associated production at the LHC. We choose the following SM input parameters~\cite{Beringer:1900zz}
\begin{eqnarray}\label{sminput}
 &&G_F=1.166379\times10^{-5}~{\rm GeV}^{-2}, ~~~~ m_H=125~{\rm GeV}, ~~~ m_Z = 91.1876~{\rm GeV}, \nno \\
 &&m_{W}=80.398~{\rm GeV},
\end{eqnarray}
and the CKM matrix is given by~\cite{Beringer:1900zz}
\begin{eqnarray}
 V_{\rm CKM}=\left(
   \begin{array}{ccc}
    0.9751 & 0.2215 & 0.0035 \\
    0.2210 & 0.9743 & 0.0410 \\
    0      & 0      & 1      \\
   \end{array}
 \right).
\end{eqnarray}
\noindent Throughout the numerical calculations, we use the MSTW2008 PDF sets and associated strong coupling constant $\as$. In order to resum all logarithmic terms $\ln\ptv/\mu_f$ to all orders, we choose the factorization scale to be $\mu_f = \ptv$~\cite{Becher:2007ty}. Besides, the hard matching scale are set as $\mu_h^2 = -M^2$ in order to contain the $\pi^2$-enhancement effects~\cite{Ahrens:2008qu}.

\subsection{Leading singular jet vetoed cross section}

\begin{figure}
\begin{center}
\includegraphics[width=0.48\textwidth]{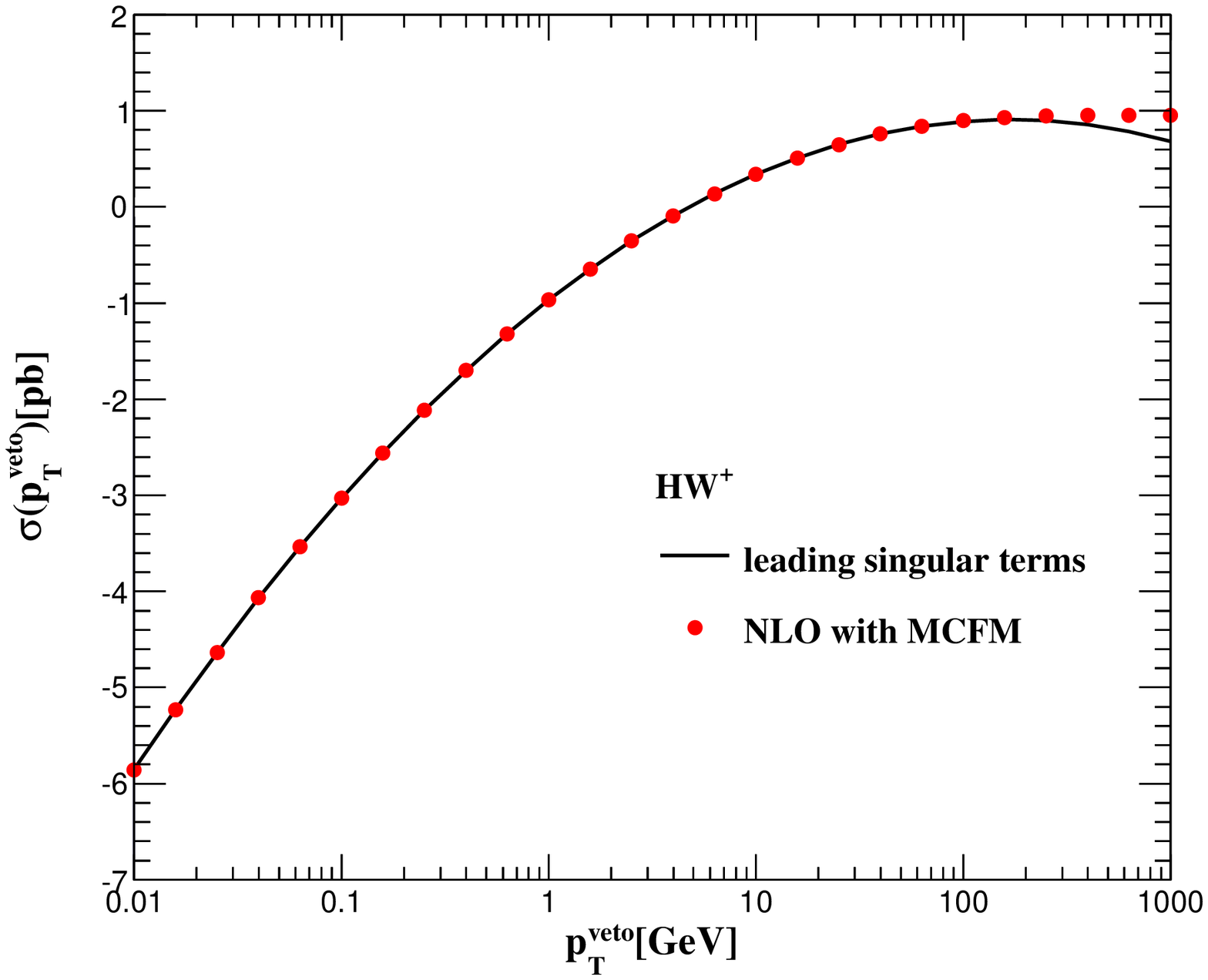}
\quad
\includegraphics[width=0.48\textwidth]{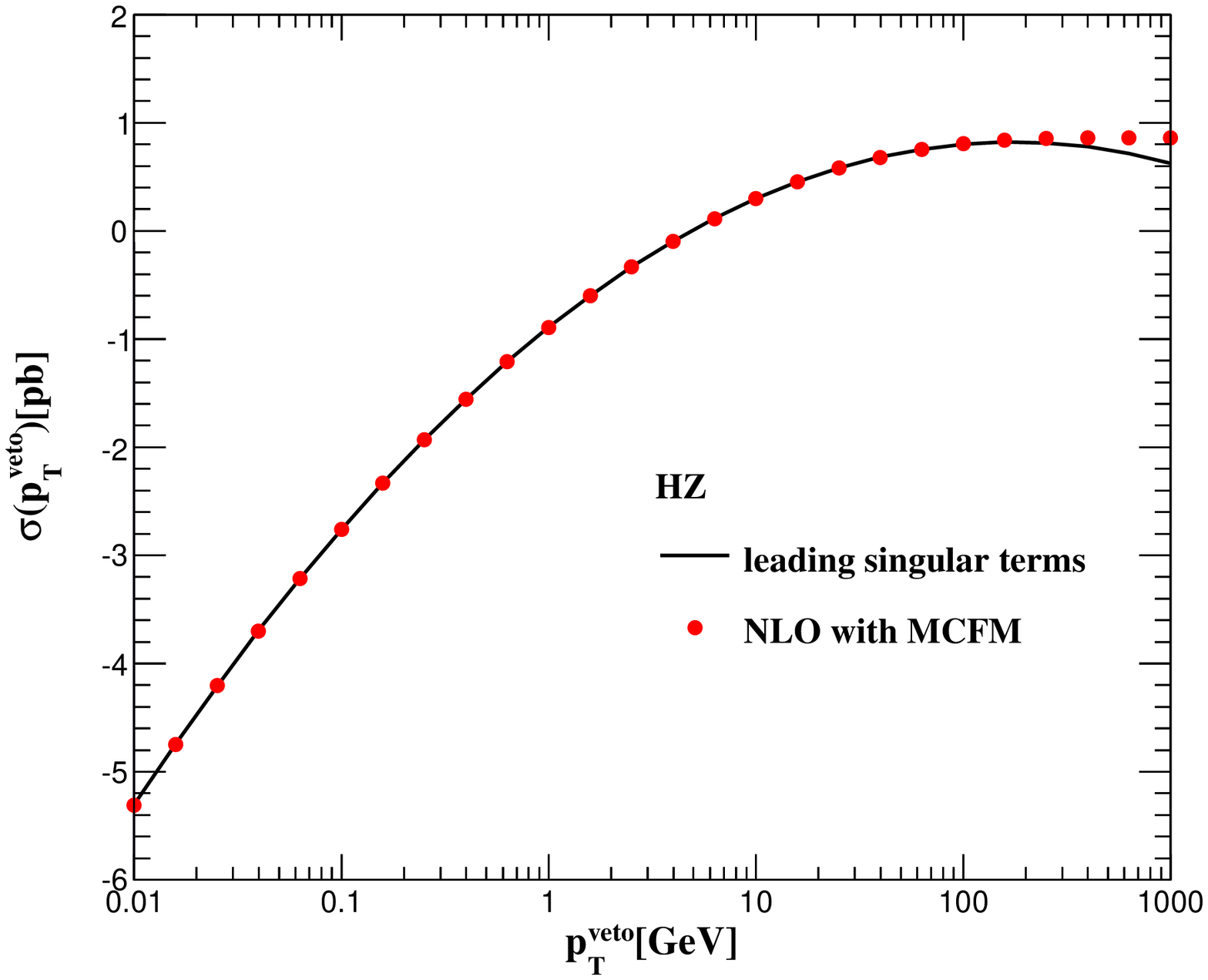}
\end{center}
\vspace{-4mm}
\caption{\label{singular_vs_NLO_plot}
Comparisons of the leading singular and the exact NLO jet vetoed cross sections for $HW^{+}$ (left panel) and $HZ$ (right panel) production at the LHC with $\sqrt{S}=14$~TeV, respectively.}
\end{figure}

For verifying the correctness of the factorization formula in Eq.~(\ref{fac_cs_3}), we expand the Eq.~(\ref{fac_cs_3}) to the leading singular terms (black solid line), and compare with the exact NLO results (red dot) calculated by modified Monte Carlo program MCFM~\cite{Campbell:2010ff} in Fig.~\ref{singular_vs_NLO_plot}. We can see that the leading singular terms of the cross section with jet veto can reproduce the exact NLO jet vetoed cross section in the small $\ptv$ region. With the increasing of $\ptv$, the difference between the leading singular and the exact NLO jet veto cross section increases.

\subsection{Scale uncertainties}

\begin{figure}[h]
\begin{center}
\includegraphics[width=0.3\textwidth]{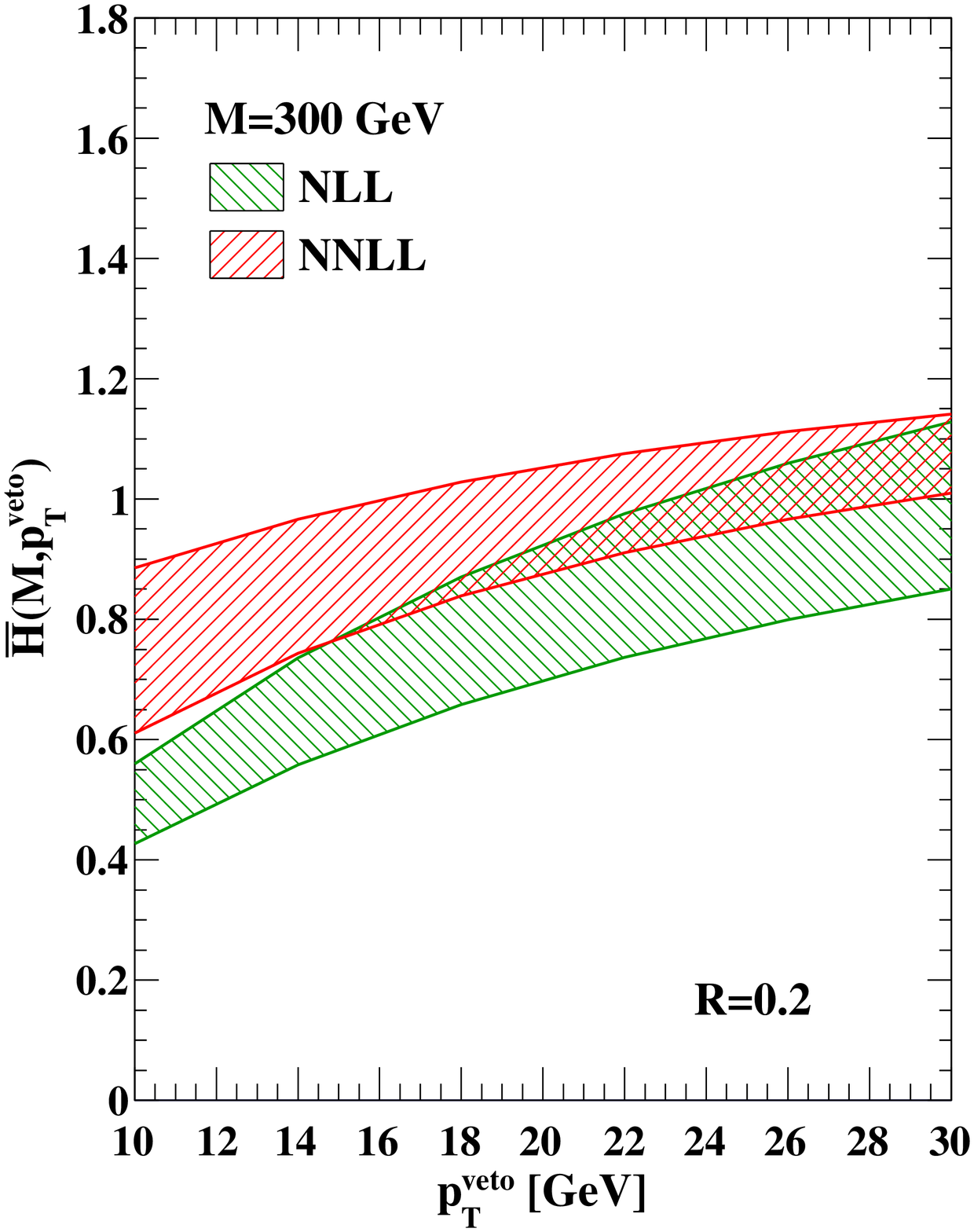}
\quad
\includegraphics[width=0.3\textwidth]{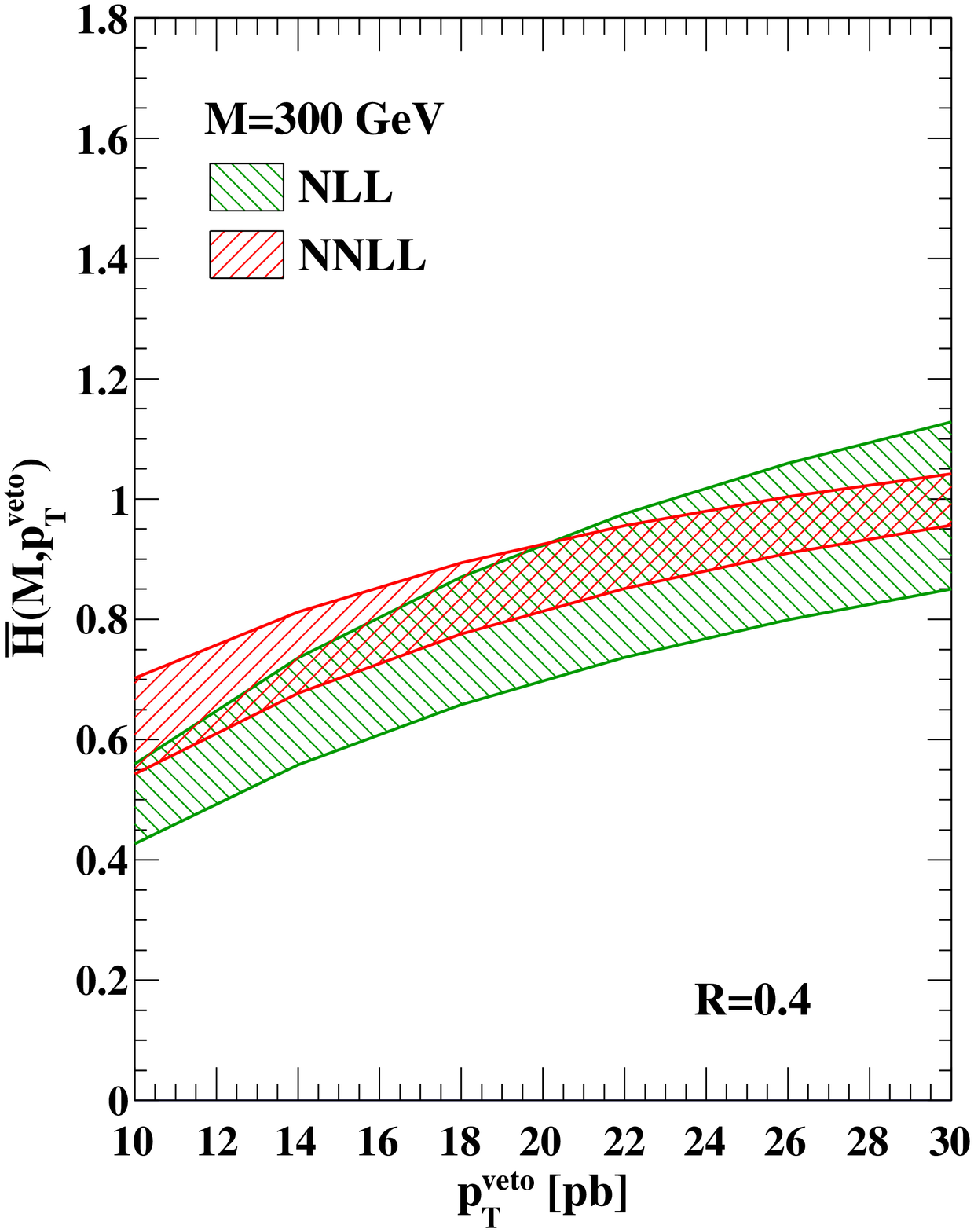}
\quad
\includegraphics[width=0.3\textwidth]{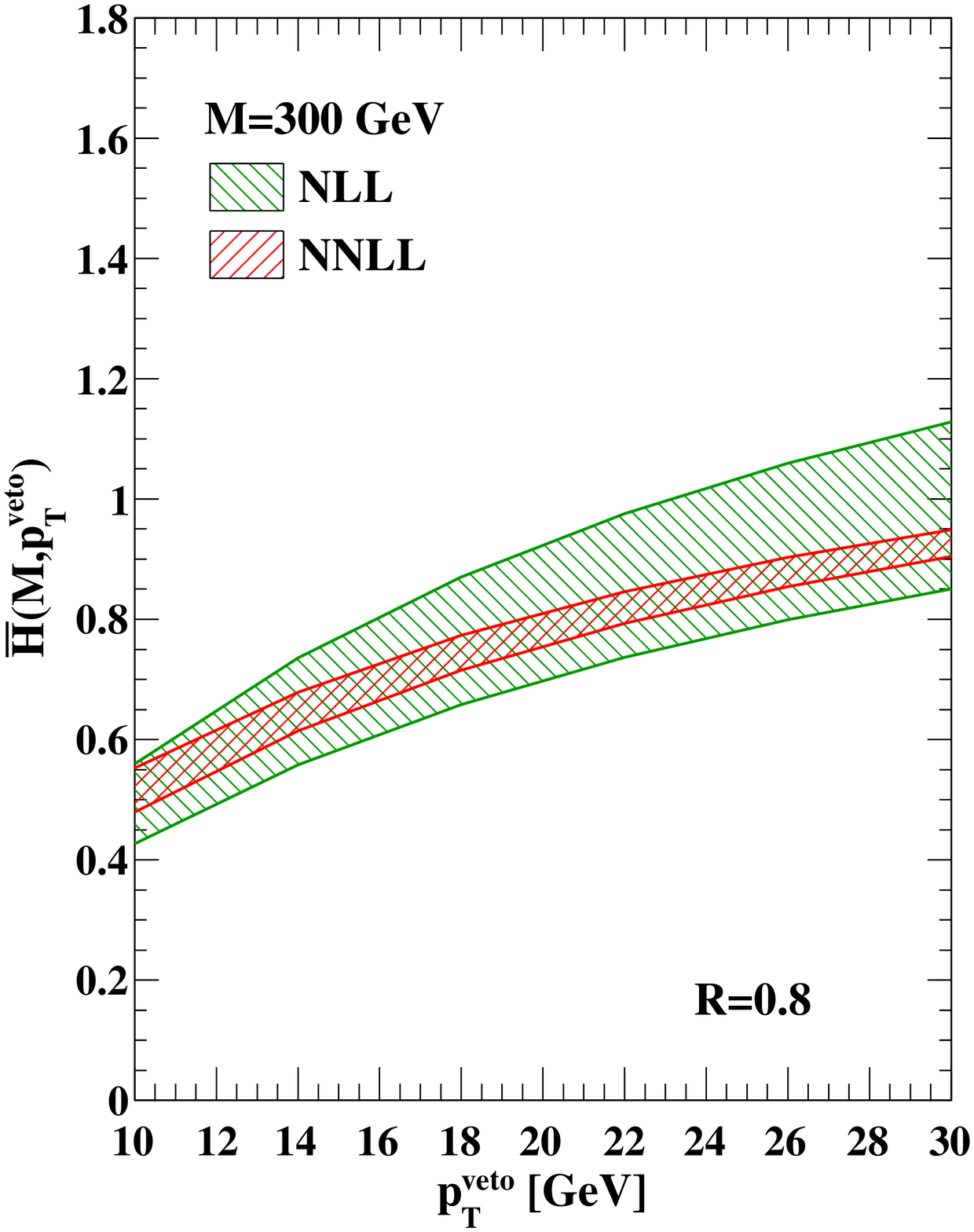}
\end{center}
\vspace{-4mm}
\caption{\label{HW_hardfun_scale_dep}
The RG invariant hard function $\overline{H}(M,\ptv)$ for three different jet radius parameter $R$, where the bands reflect the scale uncertainties, and $M=300$ GeV.}
\end{figure}

In Fig.~\ref{HW_hardfun_scale_dep} we show the scale dependence of RG invariant hard function $\overline{H}(M,\ptv)$ on $\ptv$ for three different parameters $R$, where the bands reflect the scale uncertainties by varying the scales in the range $\ptv/2<\mu_f<2\ptv$ and $M^2/4 < -\mu_h^2 < 4M^2$, respectively. In the resummation predictions these two kinds of uncertainties are added in quadrature. From Fig.~\ref{HW_hardfun_scale_dep} we can see that the NLL predictions are independent on the jet radius parameter $R$, while the NNLL predictions strongly depend on $R$. Besides, the NLL and NNLL bands overlap each other, and the scale uncertainties of NNLL results increase as $R$ decreases. When $R=0.8$, the scale uncertainties are significantly reduced from NLL level to NNLL level. And when $R=0.2$, the scale uncertainties are reduced only for large $\ptv$, and the NNLL and NLL bands overlap only for large $\ptv$ too. In the small $\ptv$ region the NNLL bands are broader than the NLL ones, and they are away from each other with the decreasing of $\ptv$.

In addition to the hard and factorization scale, another scale uncertainty coming from logarithms with collinear anomaly has also been discussed in Ref.~\cite{Becher:2013xia}, and it is shown that this uncertainty should not be included in ``collinear anomalous" formalism, although this type scale variation can be formalized in an RG framework~\cite{Chiu:2011qc,Chiu:2012ir}. Therefore, we apply the same scheme in Ref.~\cite{Becher:2013xia}, and also do not consider this kind of uncertainties in our calculations.

% After rewriting the collinear anomaly factor, we have
%\begin{eqnarray}
%  \left( \frac{M}{\ptv} \right)^{-2 F_{q\bar{q}}(\ptv,\mu)} = \left( \frac{M}{\nu} \right)^{-2 F_{q\bar{q}}(\ptv,\mu)} \left( \frac{\nu}{\ptv} \right)^{-2 F_{q\bar{q}}(\ptv,\mu)},
%\end{eqnarray}

% and the perturbative convergence becomes bad
%, and the perturbative convergence is well behaved. However,

\begin{figure}[h]
 \begin{center}
  \includegraphics[width=0.6\textwidth]{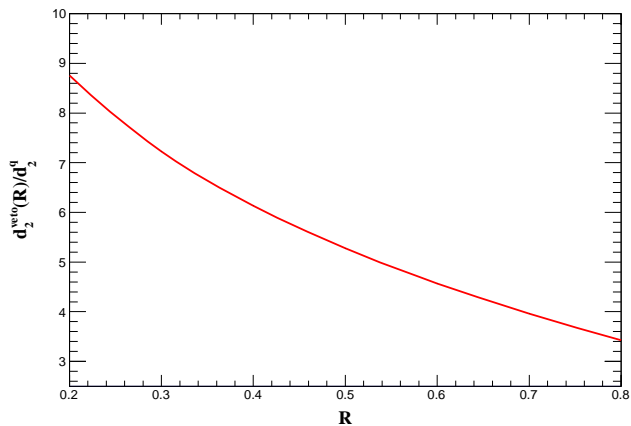}\\
  \caption{Dependence of the coefficient $d_2^{\rm veto}(R)$ on the jet radius parameter $R$, normalized to $d_2^q$.}\label{Fig:d2veto}
  \end{center}
\end{figure}

At the NNLL level the dependence of the RG invariant hard function $\overline{H}(M,\ptv)$ on the jet radius parameter $R$ is caused from the two loop anomaly coefficient $d_2^{\rm veto}(R)$. The $R$ dependence term has the form as
\begin{eqnarray}\label{d2dep}
 \exp\left[ 0.54 \frac{d_2^{\rm veto}(R)}{d_2^q} \as^2(\mu) \ln\frac{M}{\ptv} \right],
\end{eqnarray}
where $\as(\mu)$ includes the remaining scale dependence. In order to estimate the scale uncertainties induced by Eq.~(\ref{d2dep}) at the NNLL level, we show the dependence for the ratio between the coefficient $d_2^{\rm veto}(R)$ and $d_2^q$ on the jet radius parameter $R$ in Fig.~\ref{Fig:d2veto}. With the increasing of the jet radius parameter $R$ from $0.2$ to $0.8$, the coefficient $d_2^{\rm veto}(R)/d_2^q$ rapidly decrease about from $9$ to $3$ due to the existence of logarithmic terms $\ln R$ in Eq.~(\ref{fR}). Therefore, as shown in Fig.~\ref{HW_hardfun_scale_dep}, the remaining scale dependence of RG invariant hard function $\overline{H}(M,\ptv)$ increases as the parameter $R$ decreases.

\begin{figure}
\begin{center}
\includegraphics[width=0.45\textwidth]{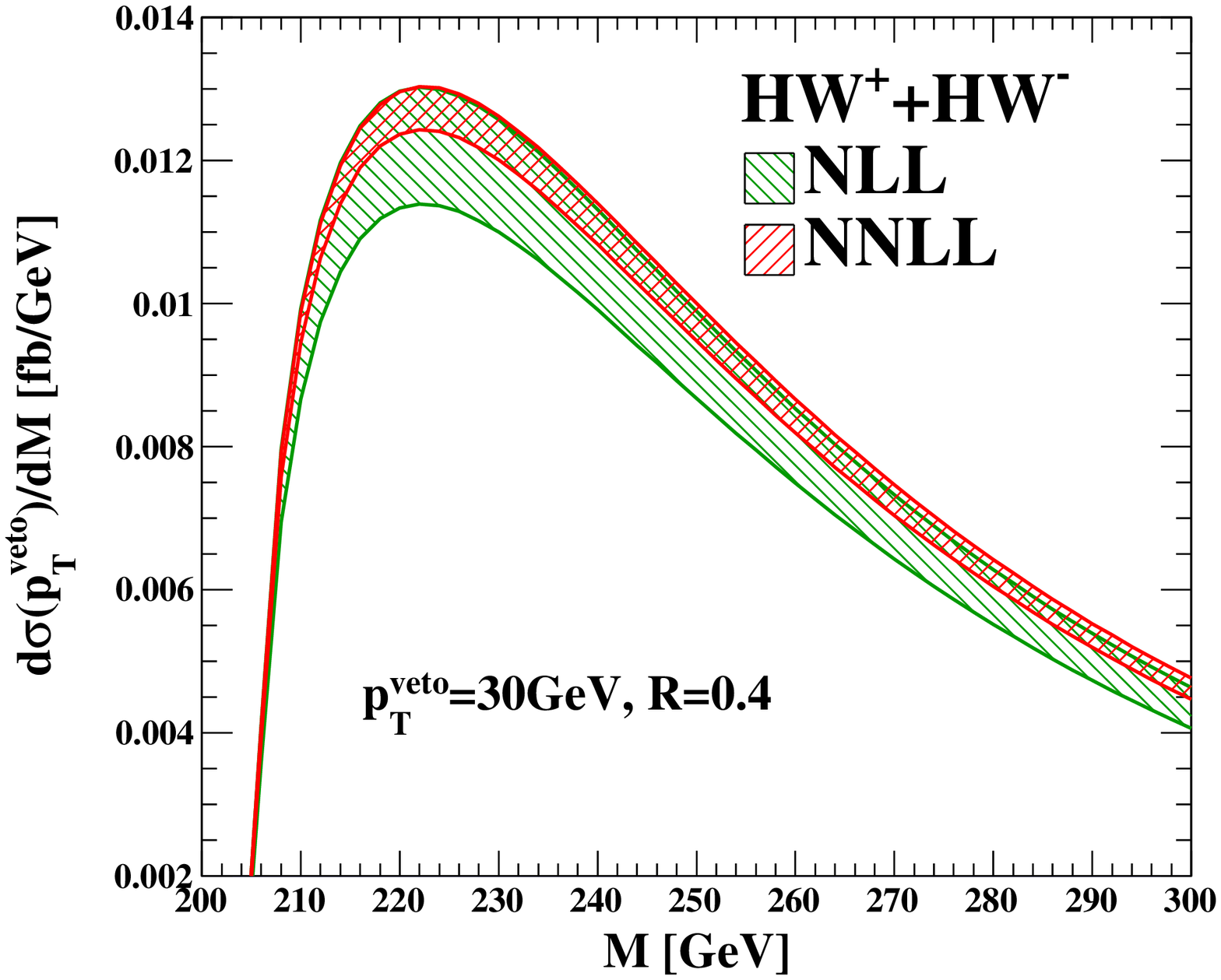}
\quad
\includegraphics[width=0.45\textwidth]{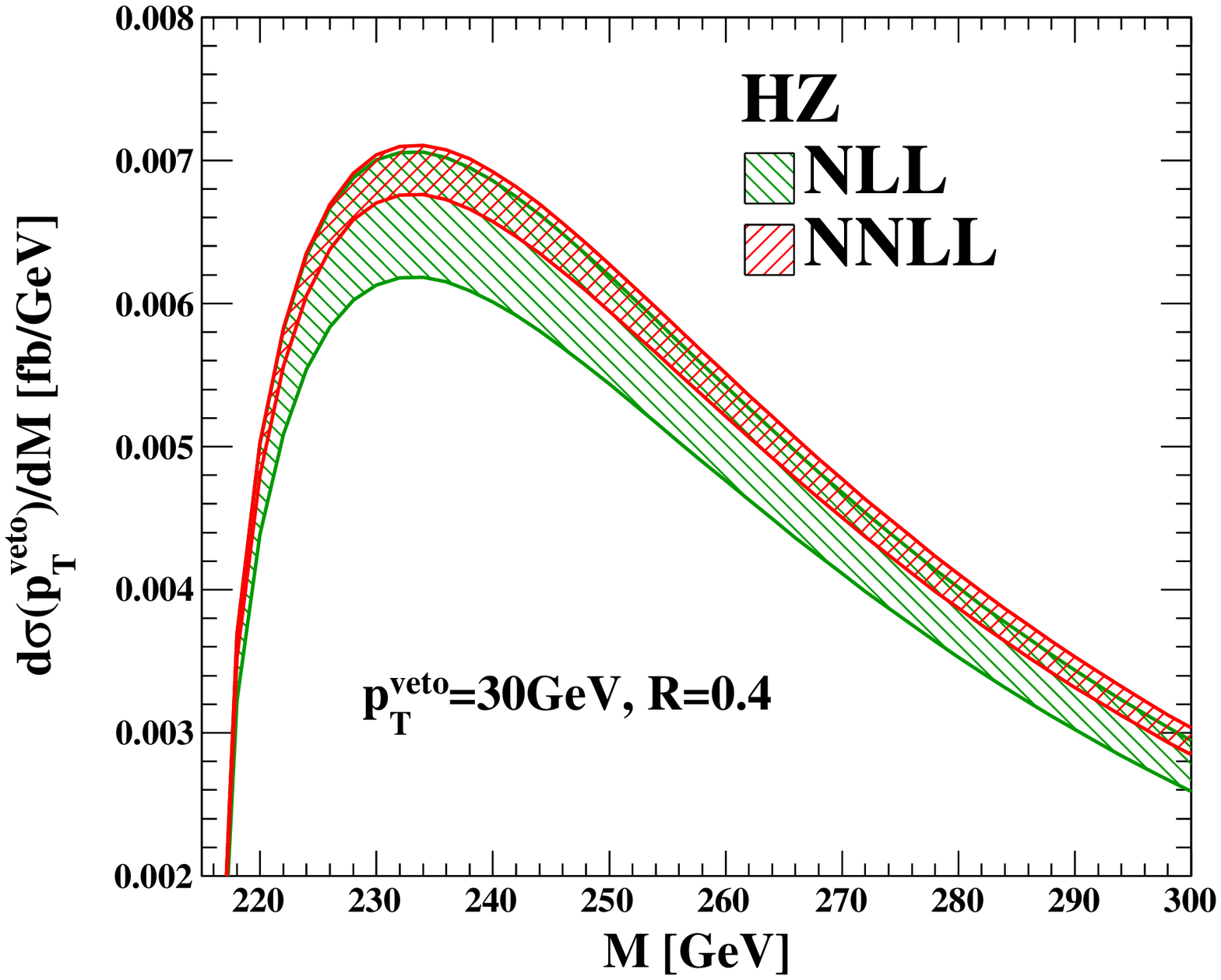}
\end{center}
\vspace{-4mm}
\caption{\label{HV_resum_InvM_sqrt_14_plot}
The NLL (green bands) and NNLL (red bands) resummed invariant mass distributions for $HW^{\pm}$ (left panel) and $HZ$ (right panel) associated production with $\ptv=20$~GeV and $R=0.4$ at the LHC with $\sqrt{S}=14$~TeV, where the bands reflect the scale uncertainties.}
\end{figure}

In Fig.~\ref{HV_resum_InvM_sqrt_14_plot} we present NLL (green bands) and NNLL (red bands) resummed predictions on the invariant mass distribution for $HV$ associated production with $\ptv=30$~GeV and $R=0.4$ at the LHC with $\sqrt{S}=14$~TeV, where the bands reflect the scale uncertainties. We use MSTW2008NLO and MSTW2008NNLO PDF sets for the NLL and NNLL results, respectively. After performing resummation, the theoretical perturbative convergence is well behaved, and the scale uncertainties are reduced from NLL level to NNLL level for all the invariant mass region.

\begin{figure}[h]
\begin{center}
\includegraphics[width=0.3\textwidth]{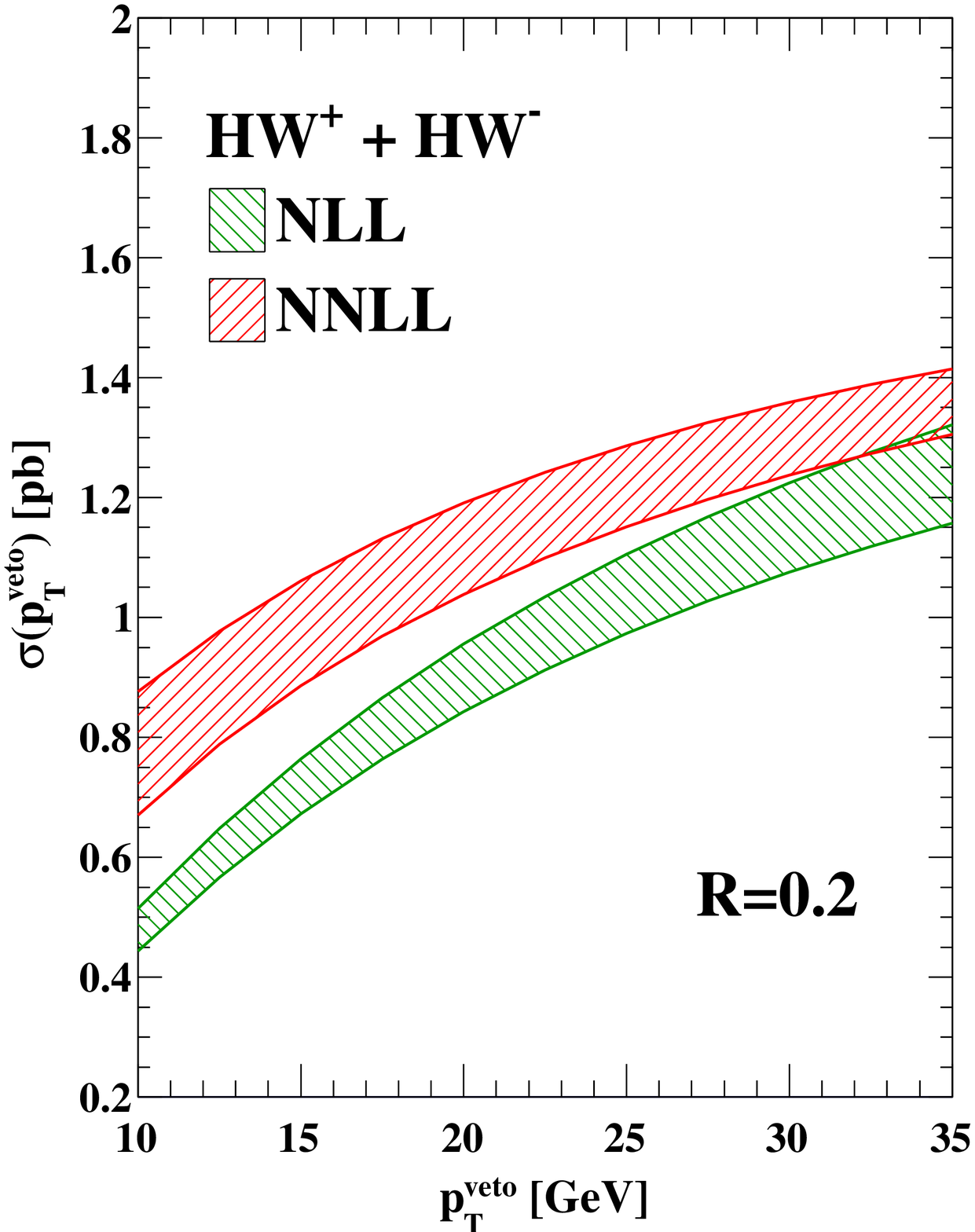}
\quad
\includegraphics[width=0.3\textwidth]{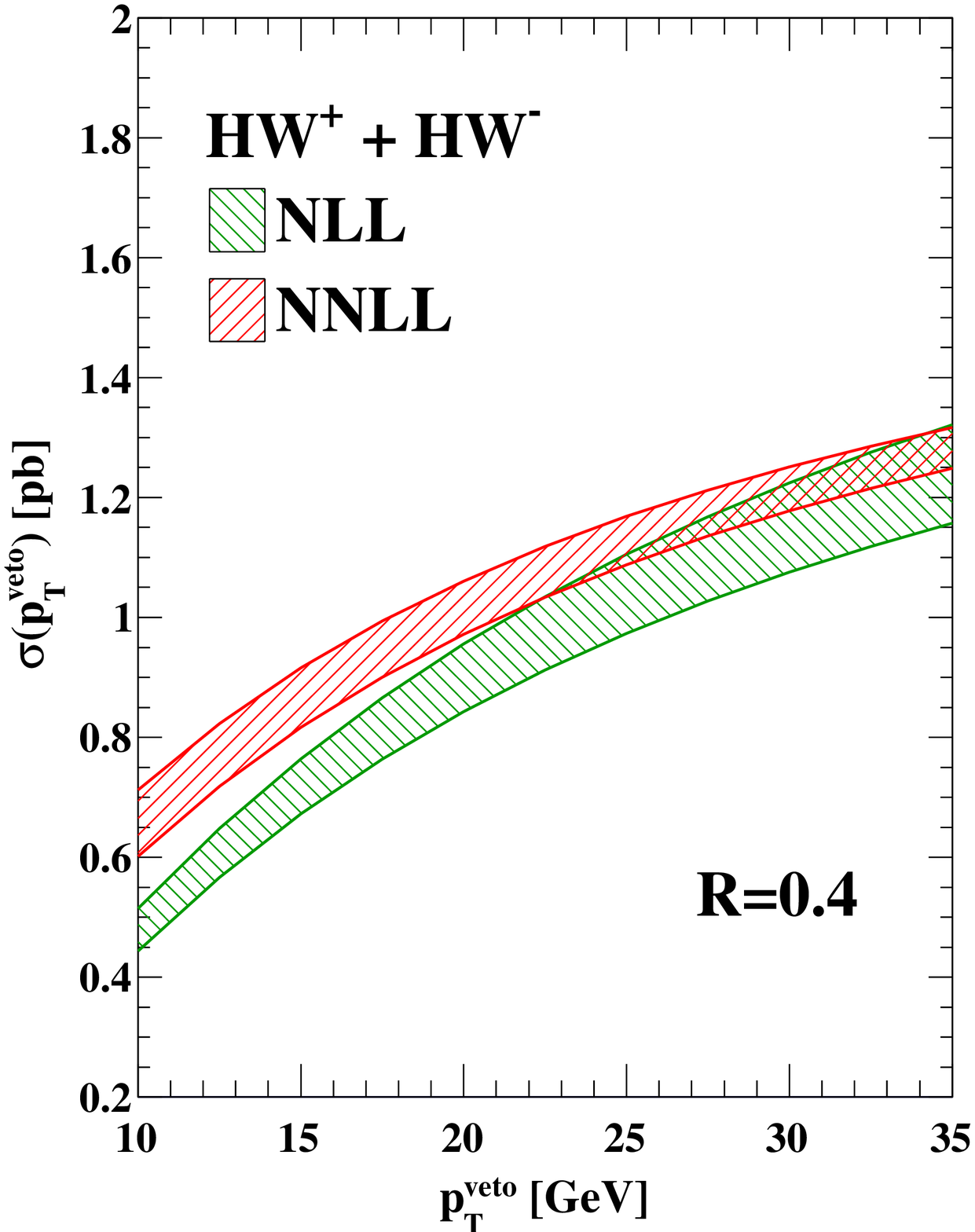}
\quad
\includegraphics[width=0.3\textwidth]{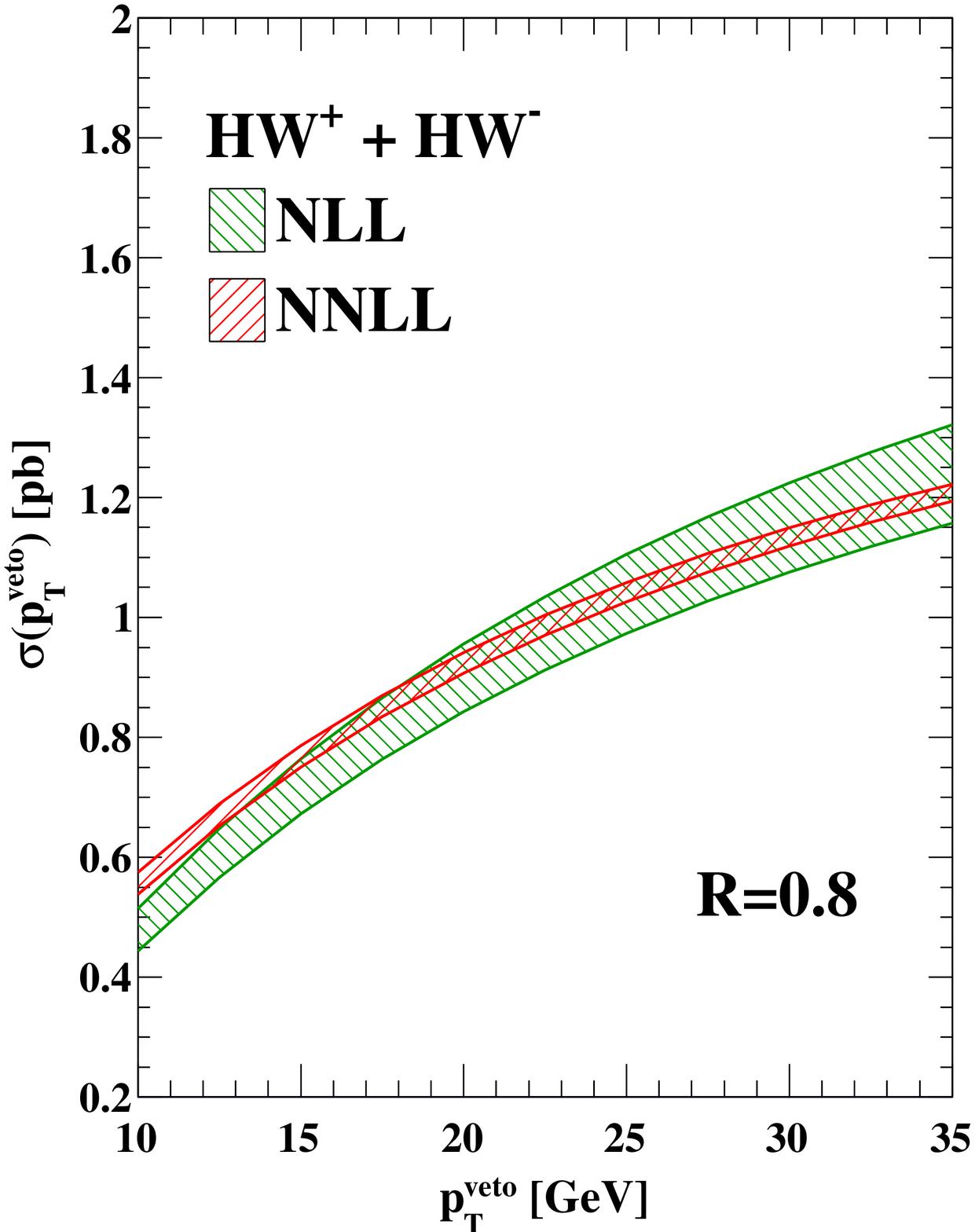}
\quad
\includegraphics[width=0.3\textwidth]{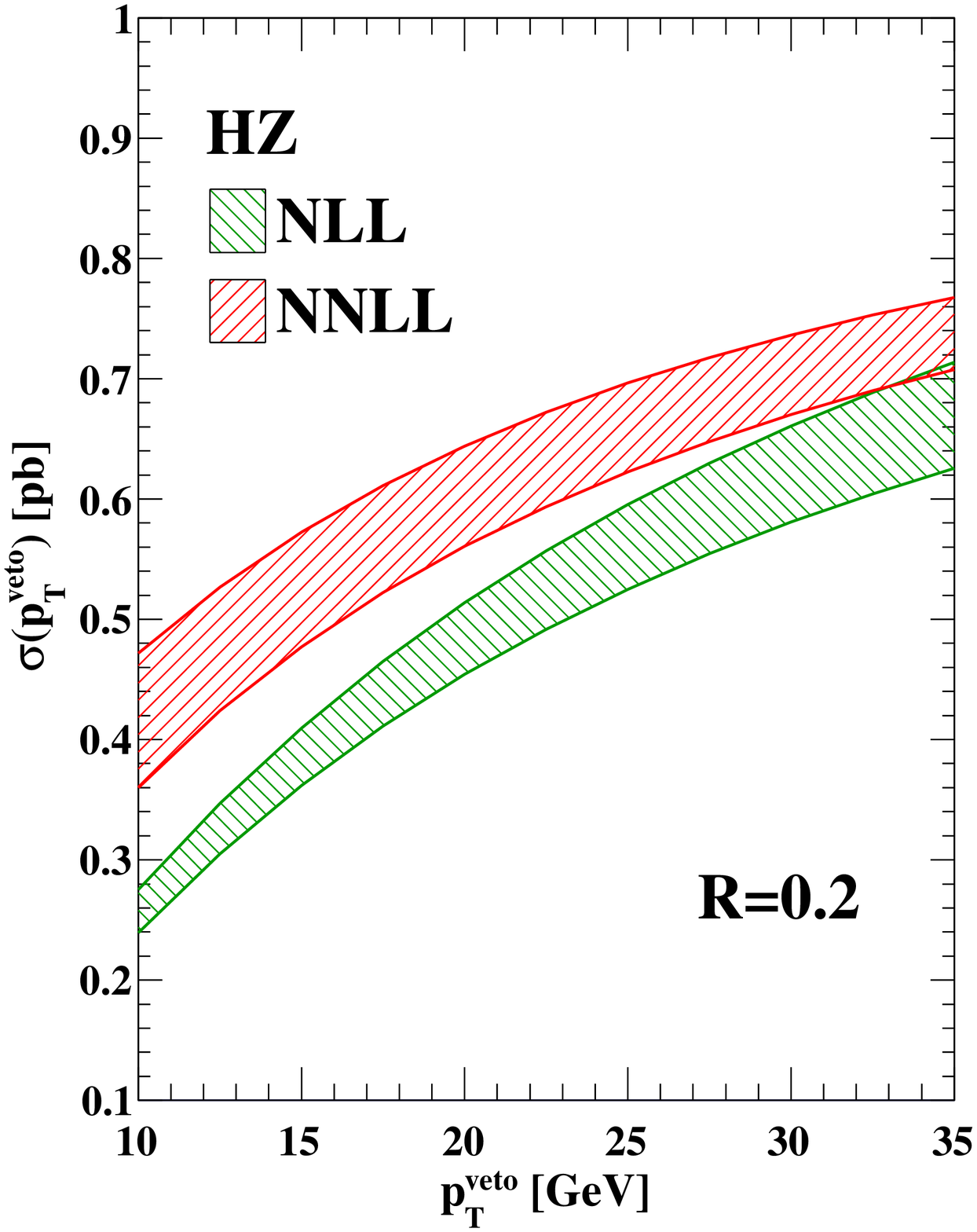}
\quad
\includegraphics[width=0.3\textwidth]{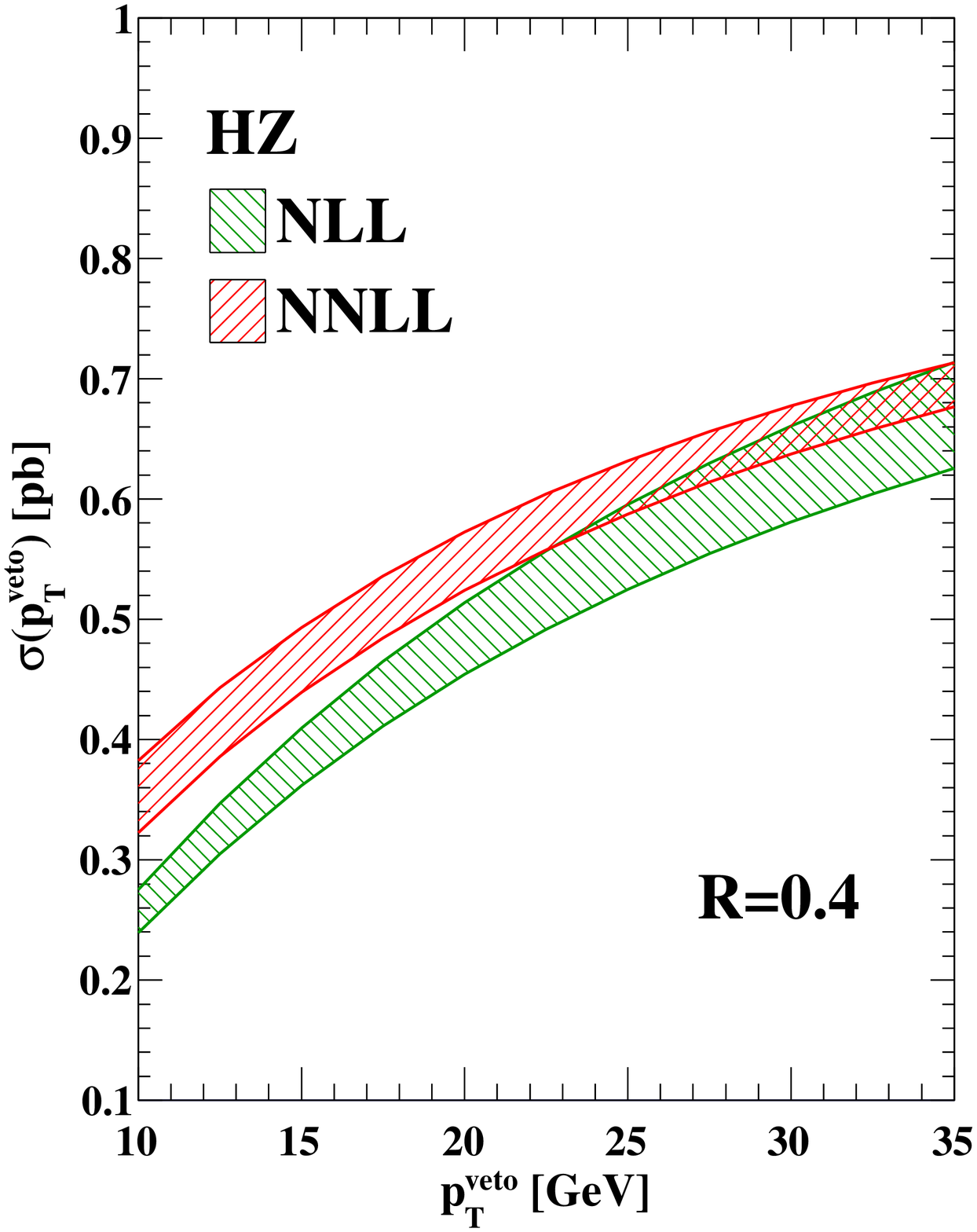}
\quad
\includegraphics[width=0.3\textwidth]{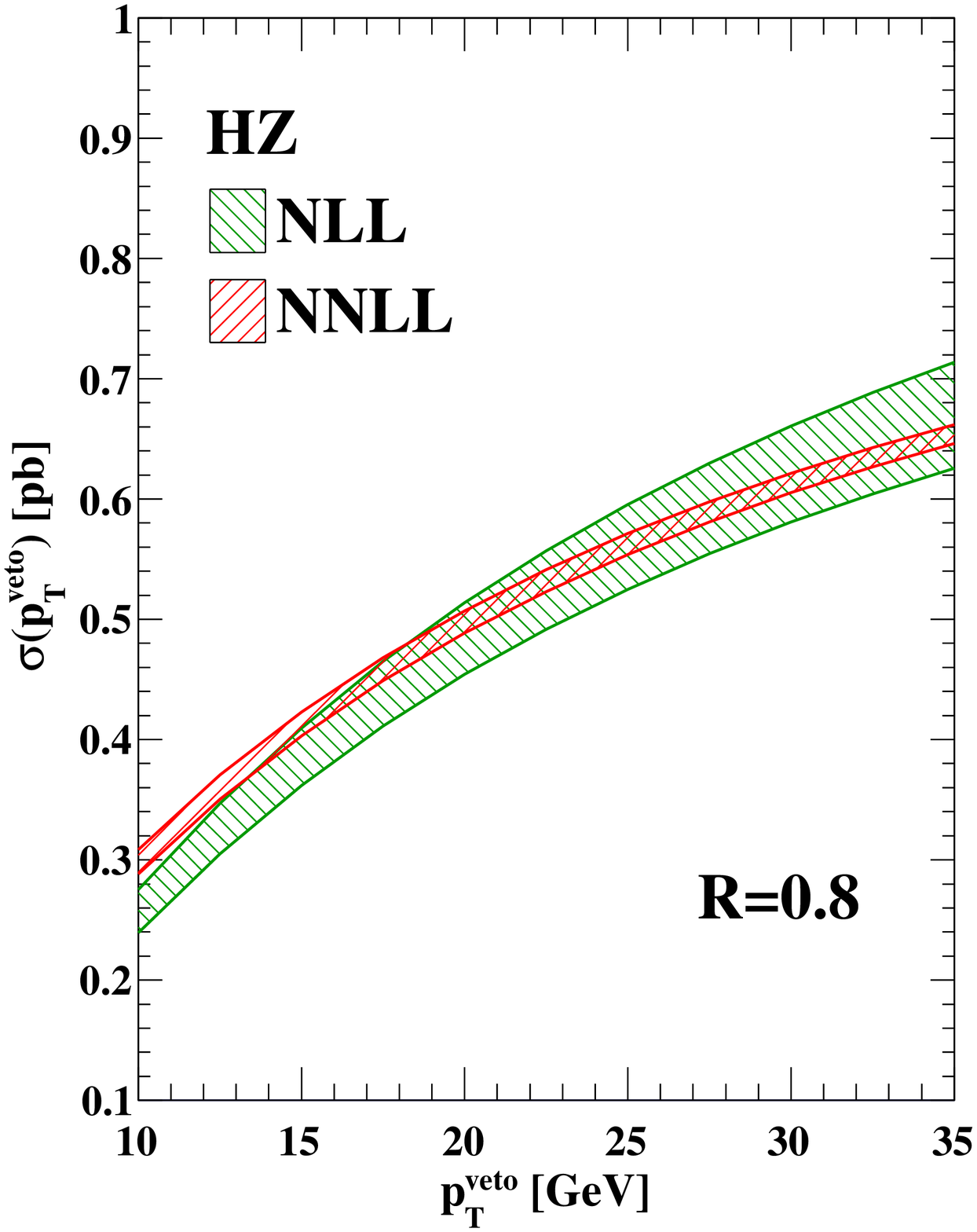}
\end{center}
\vspace{-4mm}
\caption{\label{HV_resum_veto_14}
The NLL (green bands) and NNLL (red bands) resummed jet veto cross section for $HV$ associated production at the LHC with $\sqrt{S}=14$~TeV for three different jet radius parameter $R = 0.2, 0.4$ and $0.8$, where the bands reflect the scale uncertainties.}
\end{figure}

In Fig.~\ref{HV_resum_veto_14}, we show the scale dependence of the NLL (green bands) and NNLL (red bands) resummed jet veto cross section on $\ptv$ at the LHC with $\sqrt{S}=14$~TeV for three different parameters $R = 0.2, 0.4$ and $0.8$, where the bands reflect the scale uncertainties. In the case of $HW^{\pm}$ production, the resummed jet veto cross section at the NLL level is independent on the radius parameter $R$, and the scale uncertainties are about $13\%$. Similar to the case of RG invariant hard function, with the decreasing of the parameter $R$, the scale uncertainties of NNLL results increase. When $R=0.8, 0.4$ and $0.2$, the scale uncertainties at the NNLL level are reduced to $2\%, 5\%$ and $8\%$ for $\ptv = 35$, and $7\%, 10\%$ and $17\%$ for $\ptv = 10$ GeV, respectively. Obviously, the scale uncertainties are reduced when $R=0.8$. Besides, in the large $\ptv$ region the scale uncertainties are also reduced and the NNLL and NLL bands also overlap when $R=0.2$ and $0.4$. However, in the small $\ptv$ region the NNLL uncertainties are larger than the NLL ones, and NNLL and NLL bands are away from each other. The origin of these $R$ dependence is also caused from Eq.~(\ref{d2dep}).

\subsection{RG improved phenomenology predictions at the LHC}

$HV$ associated production is an important process to study the Higgs boson at the LHC. Both of two decay modes, $h \rightarrow b\bar{b}$ and $h \rightarrow W^+ W^-$ have been searched by the ALTAS~\cite{ATLAS-CONF-2013-075,ATLAS-CONF-2013-079} and CMS~\cite{CMS-PAS-HIG-13-017,Chatrchyan:2013zna} collaborations, respectively. The results from ATLAS show that no significant excess is observed over the SM expectations, with or without a $m_H = 125$ GeV Higgs boson. And the results from CMS show that a small excess above the SM background expectation is found. Since there does not exit enough $HV$ events produced at the LHC, the corresponding jet veto studies can not be completed. With the increasing of the luminosity, $HV$ production will be more important to study the property of the SM Higgs boson, and the studies about jet veto for this process will also be attracted more attentions from experimentalists.

\begin{figure}[h]
\begin{center}
\includegraphics[width=0.45\textwidth]{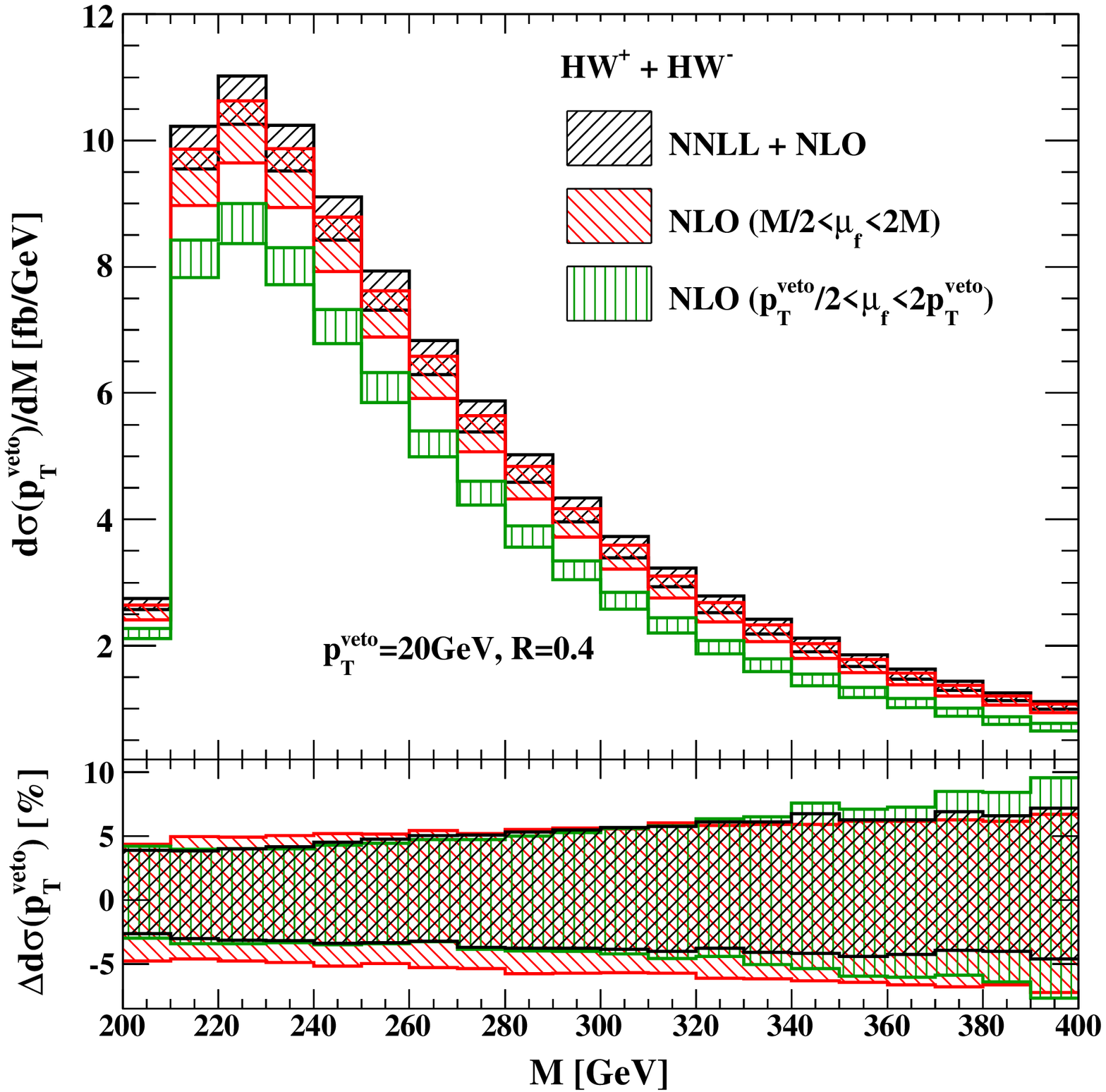}
\quad
\includegraphics[width=0.45\textwidth]{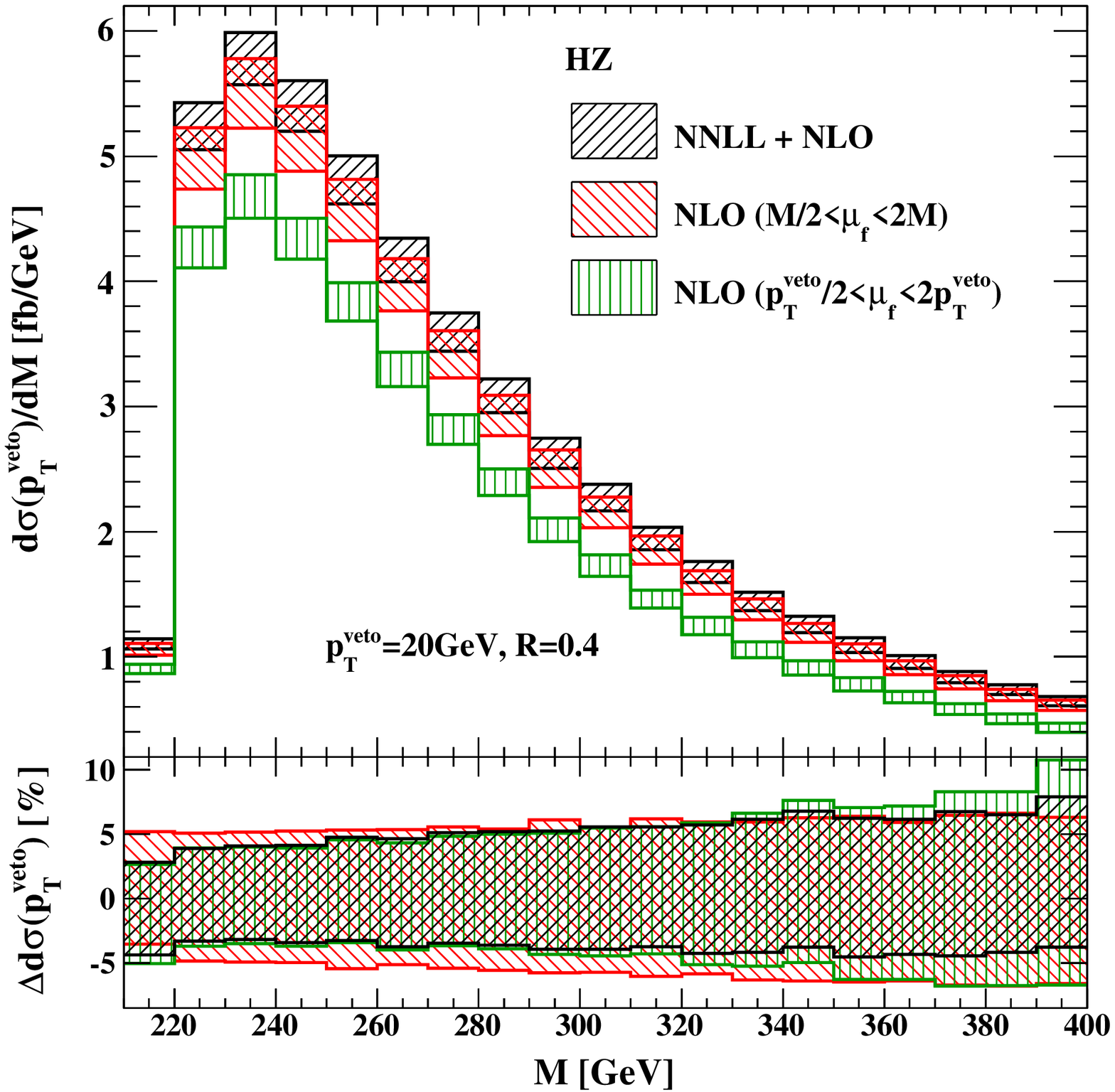}
\end{center}
\vspace{-4mm}
\caption{\label{HV_InvM_sqrt_14_plot}
The NLO and NLO+NNLL predictions for $HV$ associated production invariant mass distribution with $\ptv=20$~GeV and $R=0.4$ at the LHC with $\sqrt{S}=14$~TeV, where the bands reflect the scale uncertainties.}
\end{figure}

In Fig.~\ref{HV_InvM_sqrt_14_plot} we present the NLO+NNLL and NLO jet vetoed invariant mass distribution for $HV$ associated production at the LHC with $\sqrt{S}=14$~TeV, where $\ptv=20$~GeV and $R=0.4$ are chosen. The bands represent the scale uncertainties. We present the NLO results in two benchmark schemes, $\mu_f \sim M$ (red bands) and $\mu_f \sim \ptv$ (green bands), respectively. Compared to NLO+NNLL results (black bands), for $\mu_f \sim M$ the NLO predictions are similar to the NLO+NNLL ones, but suffer from large scale uncertainties in all the invariant mass region. However, when $\mu_f \sim \ptv$, the NLO predictions have large scale uncertainties only in the large invariant mass region, but underestimate the theoretical prediction in all the invariant mass region.

\begin{figure}[h]
\begin{center}
\includegraphics[width=0.45\textwidth]{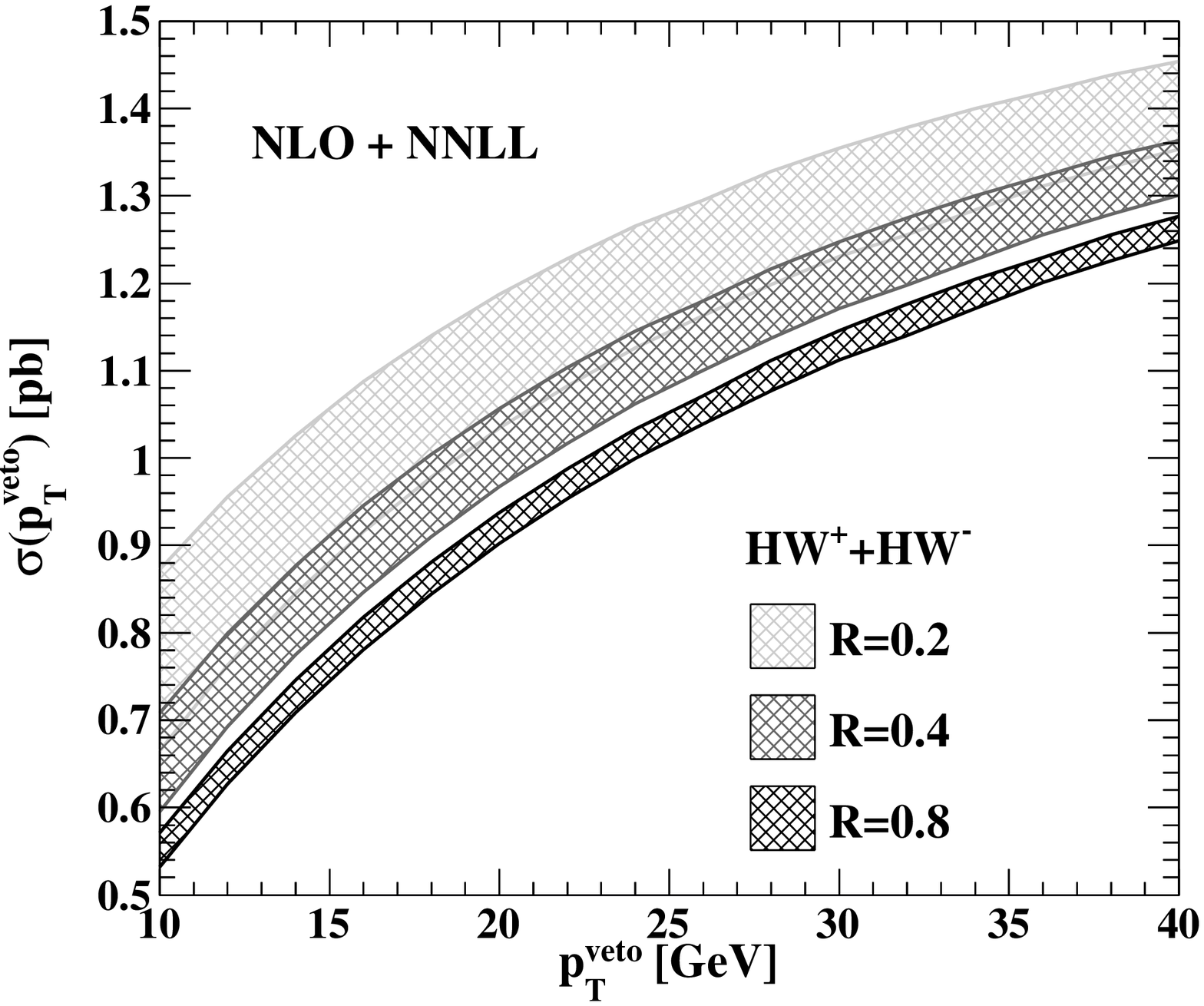}
\quad
\includegraphics[width=0.45\textwidth]{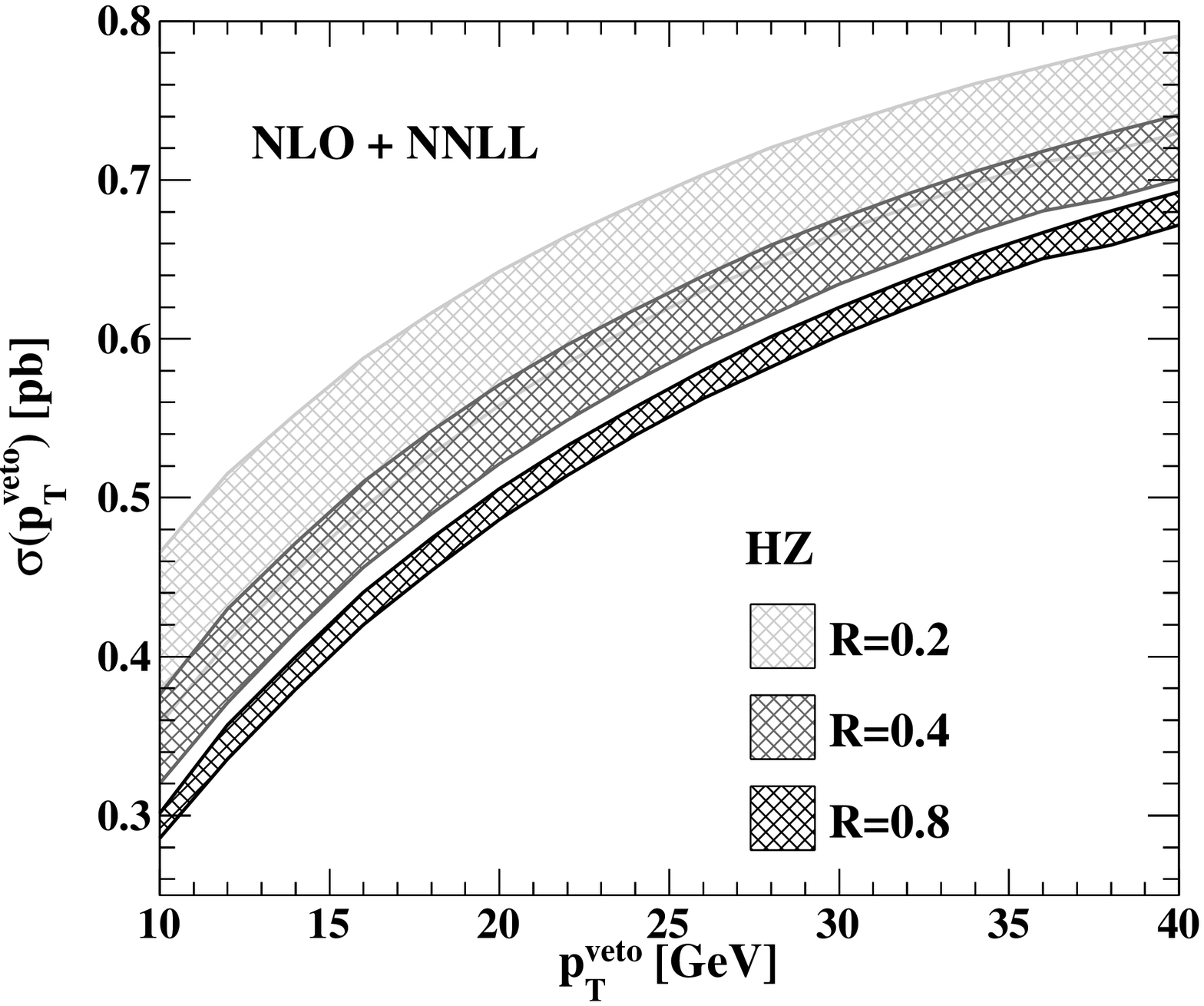}
\end{center}
\vspace{-4mm}
\caption{\label{HV_RGI_veto_14}
The NLO+NNLL predictions for $HV$ associated production cross section with a jet veto at the $14$~TeV LHC for $R = 0.2, 0.4$ and $0.8$, where the bands reflect the scale uncertainties.}
\end{figure}

After performing the integration over the invariant mass, we can get the jet vetoed cross sections. In Fig.~\ref{HV_RGI_veto_14} we present the NLO+NNLL jet vetoed cross section at the $14$~TeV LHC for $R = 0.2, 0.4$ and $0.8$, where the bands reflect the scale uncertainties. It is shown that the NLO+NNLL predictions strongly depend on the jet radius parameter $R$. With the increasing of $R$ value, the NLO+NNLL predictions decrease and the scale uncertainties reduce.

\begin{figure}[h]
\begin{center}
\includegraphics[width=0.45\textwidth]{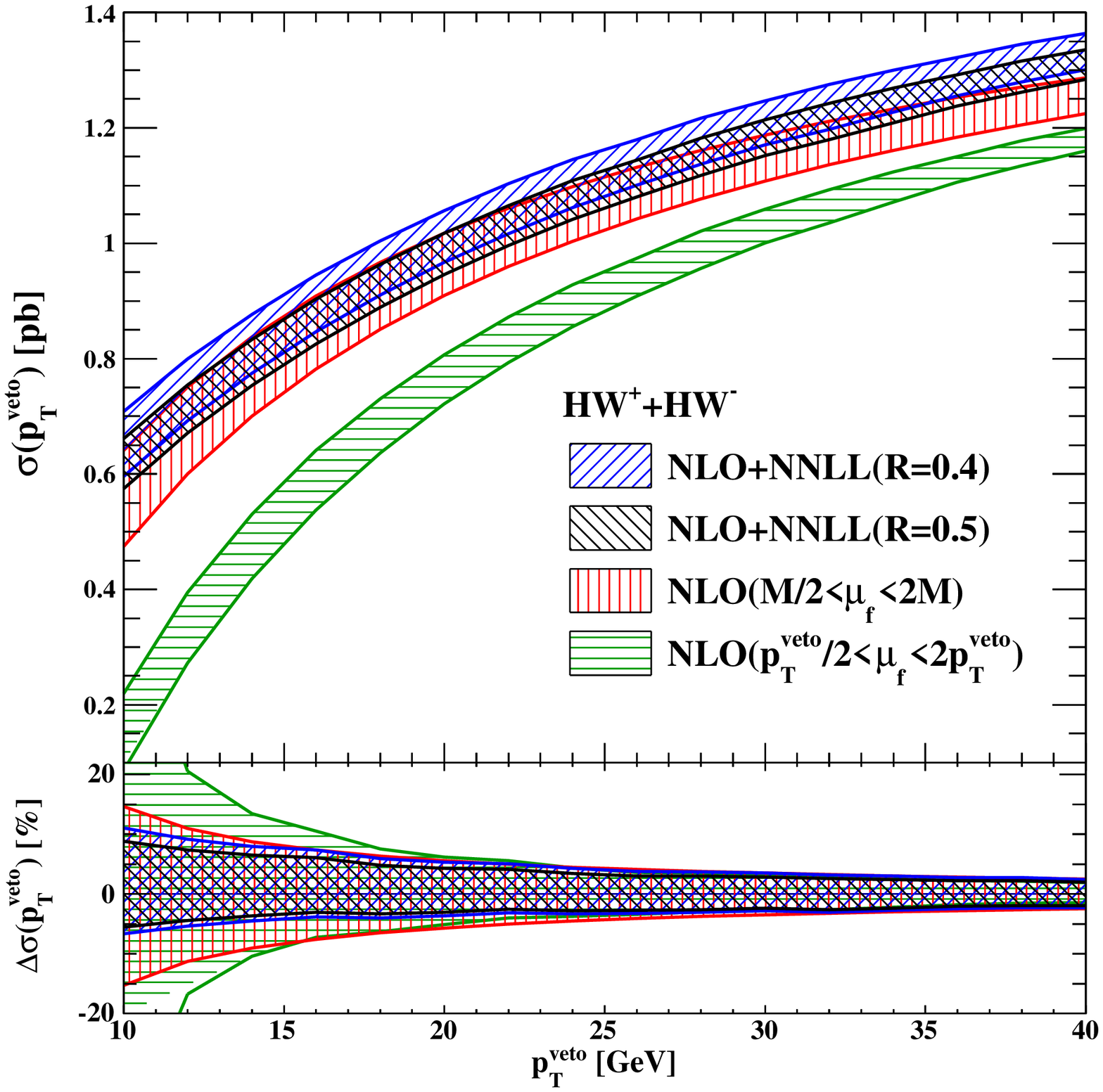}
\quad
\includegraphics[width=0.45\textwidth]{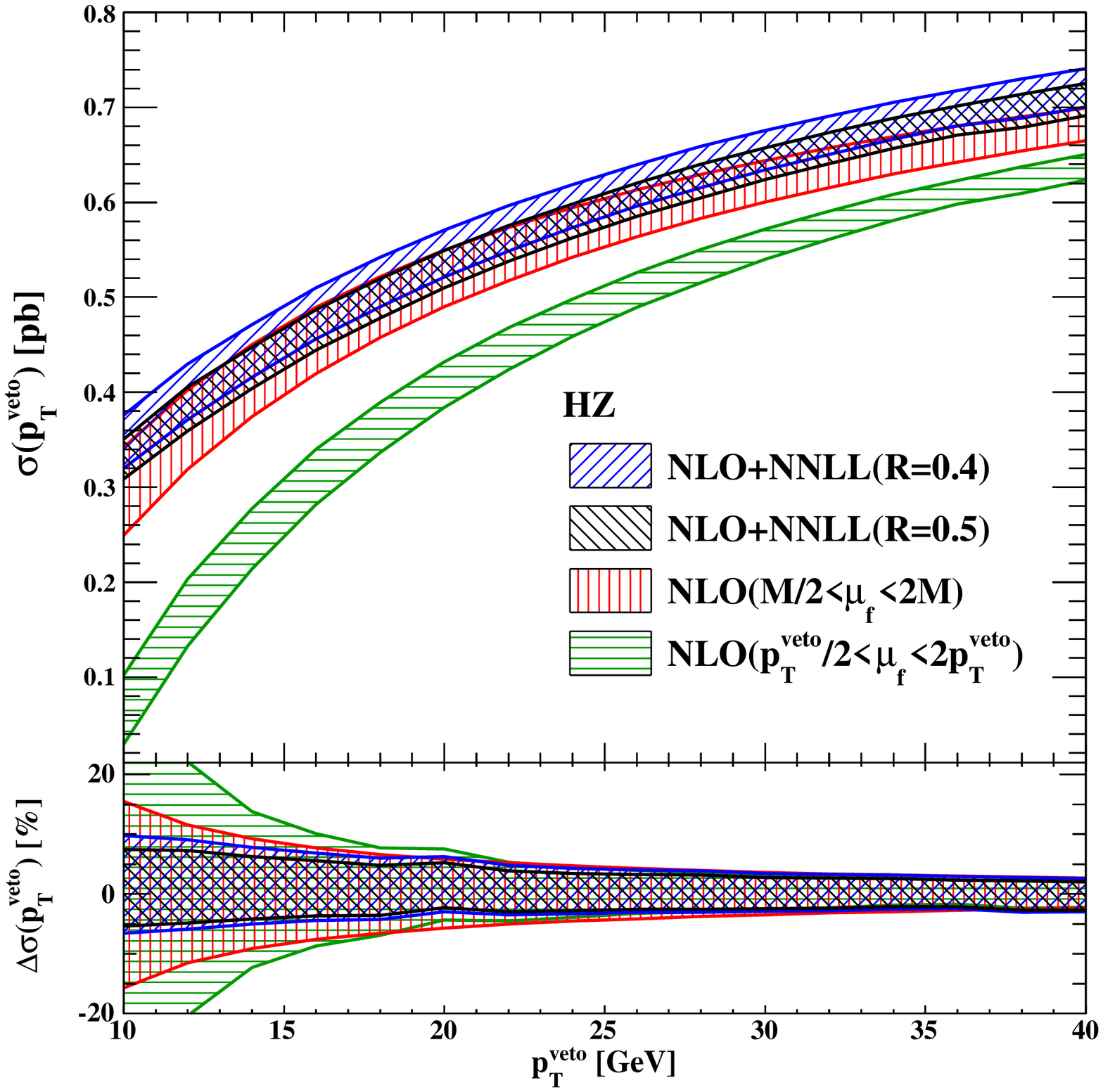}
\end{center}
\vspace{-4mm}
\caption{\label{HV_RGI_vs_NLO}
The NLO and NLO+NNLL predictions for jet vetoed cross section at the $14$~TeV LHC, where the bands reflect the scale uncertainties.}
\end{figure}

In Fig.~\ref{HV_RGI_vs_NLO}, we present the NLO and NLO+NNLL predictions for jet vetoed cross section at the $14$~TeV LHC, where the bands reflect the scale uncertainties. In the NLO+NNLL predictions the jet radius parameters $R$ are chosen as $R=0.4$ (blue bands) and $0.5$ (black bands), respectively. Besides, the NLO results are presented in two benchmark schemes, $\mu_f \sim M$ (red bands) and $\mu_f \sim \ptv$ (green bands), respectively. Obviously, the NLO results suffer from much larger scale uncertainties than the NLO+NNLL predictions in the small $\ptv$ region. Especially, when $\mu_f \sim \ptv$ is chosen, the NLO predictions break down in the small $\ptv$ region, while after including resummation effects the theoretical convergence are well behaved.

\begin{table}[h]
\begin{center}
\begin{tabular}{lccccccc}
\hline
\hline
& \multicolumn{3}{c}{$ R=0.4 $} &  & \multicolumn{3}{c}{$ R=0.5 $} \\
\cmidrule{2-4}\cmidrule{6-8}
$\ptv$ [GeV] & $20$  & $25$ & $30$ & & $20$ & $25$ & $30$ \\
\hline
$\sigma^{HW}$ [pb] & $0.92$ & $1.03$ & $1.10$ &  & $0.90$ & $1.00$ & $1.08$  \\
Scale $[\%]$ &\footnotesize $+5.1-3.7\,$ &\footnotesize $\,+4.3-3.2\,$ &\footnotesize $\,+3.5-2.8\,$ & & \footnotesize $+4.1-3.1\,$ &\footnotesize $\,+3.5-2.7\,$ &\footnotesize $\,+2.8-2.3\,$  \\
PDF $[\%]$ &\footnotesize $+4.0-3.6\,$ &\footnotesize $\,+3.9-3.5\,$ &\footnotesize $\,+3.8-3.4\,$ & & \footnotesize $+4.0-3.6\,$ &\footnotesize $\,+3.9-3.5\,$ &\footnotesize $\,+3.8-3.4\,$ \\
\hline
$\sigma^{HZ}$ [pb] & $0.498$ & $0.554$ & $0.598$ &  & $0.484$ & $0.541$ & $0.585$  \\
Scale $[\%]$ &\footnotesize $+5.5-3.9\,$ &\footnotesize $\,+4.3-3.4\,$ &\footnotesize $\,+3.6-2.9\,$ & & \footnotesize $+4.5-3.3\,$ &\footnotesize $\,+3.5-2.9\,$ &\footnotesize $\,+2.9-2.5\,$  \\
PDF $[\%]$ &\footnotesize $+4.0-3.5\,$ &\footnotesize $\,+3.9-3.3\,$ &\footnotesize $\,+3.8-3.3\,$ & & \footnotesize $+4.0-3.5\,$ &\footnotesize $\,+3.9-3.3\,$ &\footnotesize $\,+3.8-3.3\,$ \\
\hline\hline
\end{tabular}
\caption{\label{tab:jet_veto_cs_sqrt_13}
The jet vetoed cross section at the $13$~TeV LHC with jet radius parameter $R = 0.4$ and $0.5$, respectively.}
\end{center}
\end{table}

\begin{table}[h]
\begin{center}
\begin{tabular}{lccccccc}
\hline
\hline
& \multicolumn{3}{c}{$ R=0.4 $} &  & \multicolumn{3}{c}{$ R=0.5 $} \\
\cmidrule{2-4}\cmidrule{6-8}
$\ptv$ [GeV] & $20$  & $25$ & $30$ & & $20$ & $25$ & $30$ \\
\hline
$\sigma^{HW}$ [pb] & $1.00$ & $1.12$ & $1.20$ &  & $0.98$ & $1.08$ & $1.17$  \\
Scale $[\%]$ &\footnotesize $+5.3-3.6\,$ &\footnotesize $\,+4.4-3.0\,$ &\footnotesize $\,+3.5-2.8\,$ & & \footnotesize $+4.3-3.0\,$ &\footnotesize $\,+3.6-2.5\,$ &\footnotesize $\,+2.9-2.4\,$  \\
PDF $[\%]$ &\footnotesize $+3.9-3.5\,$ &\footnotesize $\,+3.9-3.5\,$ &\footnotesize $\,+3.8-3.4\,$ & & \footnotesize $+3.9-3.5\,$ &\footnotesize $\,+3.9-3.5\,$ &\footnotesize $\,+3.8-3.4\,$ \\
\hline
$\sigma^{HZ}$ [pb] & $0.537$ & $0.604$ & $0.653$ &  & $0.522$ & $0.591$ & $0.640$  \\
Scale $[\%]$ &\footnotesize $+6.3-2.9\,$ &\footnotesize $\,+4.0-3.2\,$ &\footnotesize $\,+3.4-2.9\,$ & & \footnotesize $+5.3-2.3\,$ &\footnotesize $\,+3.2-2.7\,$ &\footnotesize $\,+2.8-2.5\,$  \\
PDF $[\%]$ &\footnotesize $+4.0-3.4\,$ &\footnotesize $\,+3.8-3.3\,$ &\footnotesize $\,+3.7-3.2\,$ & & \footnotesize $+4.0-3.4\,$ &\footnotesize $\,+3.8-3.3\,$ &\footnotesize $\,+3.7-3.2\,$ \\
\hline\hline
\end{tabular}
\caption{\label{tab:jet_veto_cs_sqrt_14}
The jet vetoed cross section at the $14$~TeV LHC with jet radius parameter $R = 0.4$ and $0.5$, respectively.}
\end{center}
\end{table}

In Tab.~\ref{tab:jet_veto_cs_sqrt_13} and Tab.~\ref{tab:jet_veto_cs_sqrt_14} we list the NLO+NNLL jet vetoed cross section at the LHC with $\sqrt{S}=13$ and $14$ TeV, respectively. Here, besides scale uncertainties are taken into account, to estimate the PDF uncertainties, we use the MSTW2008 $90\%$ C.L. PDF sets~\cite{Martin:2009bu}, which are known to provide very close results to the PDF4LHC working group recommendation for the envelop prescription~\cite{Botje:2011sn}. Tab.~\ref{tab:jet_veto_cs_sqrt_13} and Tab.~\ref{tab:jet_veto_cs_sqrt_14} show that the scale and PDF uncertainties are almost same order. Moreover, with the increasing of the $\ptv$ and $R$, the scale uncertainties decrease, while the PDF uncertainties almost do not change.

\section{Conclusion}\label{s5}
We have studied the resummation effects for the $HV$ associated production at the LHC with a jet veto in SCET using ``collinear anomalous" formalism. We calculate the jet vetoed invariant mass distribution and the cross section for this process at Next-to-Next-to-Leading-Logarithmic level, which are matched to the QCD Next-to-Leading Order results, and compare the differences of the resummation effects with different jet veto $\ptv$ and jet radius $R$. Our results show that both resummation enhancement effects and the scale uncertainties decrease with the increasing of jet veto $\ptv$ and jet radius $R$, respectively. When $\ptv=25$ GeV and $R=0.4~(0.5)$, the resummation effects reduce the scale uncertainties of the Next-to-Leading Order jet vetoed cross sections to about $7\%~(6\%)$, which lead to increased confidence on the theoretical predictions. Besides, after including resummation effects, the PDF uncertainties of jet vetoed cross section are about $7\%$. Our results can help to precisely study the physical property of the SM Higgs boson through Higgs and vector boson associated production at the LHC in the future.

\begin{acknowledgments}
We would like to thank Hua Xing Zhu for helpful discussions. This work was supported in part by the National Natural Science Foundation of China under Grants NO.~11375013 and NO.~11135003.
\end{acknowledgments}

\appendix
%%%%%%%%%%%%%%%%%%%%%%%%%%%% appendix A %%%%%%%%%%%%%%%%%%%%%%%
\section{Calculation of beam functions}\label{a1}
\begin{figure}[h]
\begin{center}
\includegraphics[width=0.8\textwidth]{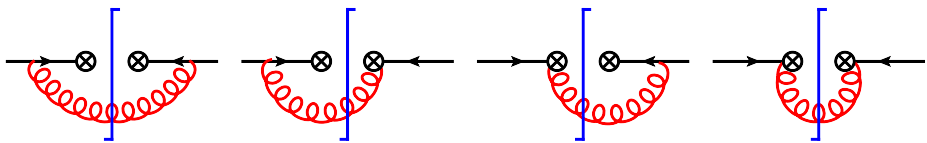}
\end{center}
\vspace{-4mm}
\caption{\label{beamfun}
Feynman diagrams contribution to the NLO beam function $\mathcal{I}_{q \leftarrow q}$.}
\end{figure}

In this appendix we show the details of calculating the beam functions. At the NLO, the beam functions receive the contributions from the diagrams shown in Fig.~\ref{beamfun} and we have the sum of these diagrams,
\begin{align}
 \mathcal{I}^{\, (1), {\rm bare}}_{q \leftarrow q}(z,\ptv,\mu)  = \, & g_s^2 C_F \mu^{2\epsilon} \int \frac{d^D k}{(2\pi)^{D-1}} \left( \frac{\nu}{k^+} \right)^\alpha \delta(k^2)\theta(k^0)\delta(k^- - (1-z) p^-)  \nno \\
 & \times \theta(\ptv - k_T) \frac{k^-}{k_T^2}\left[ (D-2)(1-z) + \frac{4z}{1-z} \right],
\end{align}
where we have suppressed the $\overline{\rm MS}$ factor $(e^{\gamma_{\rm E}}/4\pi)^\epsilon$ and the analytic regularization method of Ref.~\cite{Becher:2011dz} is used. The integration measure can be written as
\begin{align}
 d^D k \delta(k^2)\theta(k^0)\delta(k^- - (1-z) p^-) \theta(\ptv - k_T) = & \nno \\
 &  \hspace{-8em} \frac{1}{2p^-} \frac{1}{1-z} \frac{2\pi^{1/2-\epsilon}}{\Gamma(1/2-\epsilon)} \int_0^{\ptv} d k_T d\theta k_T^{1-2\epsilon} \sin^{-2\epsilon} \theta.
\end{align}
Thus, we have bare $\mathcal{I}_{q \leftarrow q}$ up to NLO,
\begin{align}
 \mathcal{I}_{q \leftarrow q}^{\rm bare}(z,\ptv,\mu) =& \, \delta(1-z) -\frac{C_F \as}{2\pi} \Bigg\{ \delta(1-z)\left( -\frac{2}{\epsilon^2} + L_\perp^2 + \frac{\pi^2}{6} \right)  \nno \\
 & \hspace{-5em}  + \left( \frac{1}{\epsilon} + L_\perp \right) \left[ \left( \frac{2}{\alpha} - 2 \ln\frac{\mu^2}{\nu p_1^-} \right) \delta(1-z) +  \frac{2}{(1-z)_+} - z - 1 \right] - (1-z) \Bigg\}.
\end{align}
Similarly, the bare $\mathcal{I}_{\bar{q} \leftarrow \bar{q}}$ is given by
\begin{align}\label{beama}
 \mathcal{I}^{{\rm bare}}_{\bar{q} \leftarrow \bar{q}}(z,\ptv,\mu) = & \, \delta(1-z)  -\frac{C_F \as}{2\pi} \Bigg\{ \left( \frac{1}{\epsilon} + L_\perp \right) \Bigg[ \left(- \frac{2}{\alpha} - 2\ln\frac{\nu}{p_2^+} \right) \delta(1-z) \nno \\
  & +  \frac{2}{(1-z)_+} - z - 1 \Bigg] - (1-z) \Bigg\}.
\end{align}
The product of two beam functions is independent on the regulator $\alpha$ and well defined in the general dimensional regularization.
\begin{figure}[h]
\begin{center}
\includegraphics[width=0.25\textwidth]{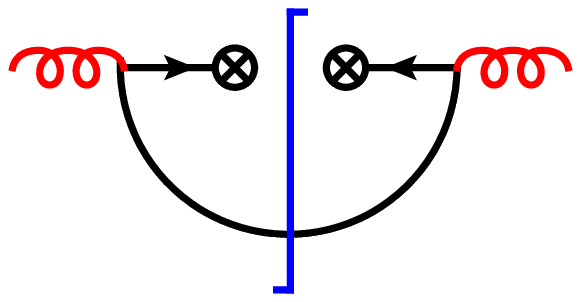}
\end{center}
\vspace{-4mm}
\caption{\label{beamfun_2}
Feynman diagram contributions to the NLO beam function $\mathcal{I}_{q \leftarrow g}$.}
\end{figure}

The evaluation of the beam function $\mathcal{I}_{q \leftarrow g}$ is independent on the regulator $\alpha$, and the corresponding Feynman diagram is shown in Fig.~\ref{beamfun_2}. After performing analytical calculation, we have
\begin{align}\label{beamb}
 \mathcal{I}^{{\rm bare}}_{\bar{q} \leftarrow g}(z,\ptv,\mu) = - \frac{T_F \as}{2\pi}\left\{ \left( \frac{1}{\epsilon} + L_\perp \right) \left[ z^2 + (1-z)^2 \right] - 2 z(1-z) \right\}.
\end{align}
By means of Eq.(\ref{beama}) and (\ref{beamb}), after $\overline{\rm MS}$ subtraction we can exact the coefficients $d_1^{\rm veto}(R)$, $R_{q \leftarrow q}(z)$ and $R_{q \leftarrow g}(z)$, directly.

\bibliographystyle{JHEP}
\bibliography{hv_veto}

\begin{thebibliography}{10}

\bibitem{Aad:2012tfa}
Georges Aad et~al.
\newblock {Observation of a new particle in the search for the Standard Model
  Higgs boson with the ATLAS detector at the LHC}.
\newblock {\em Phys.Lett.}, B716:1--29, 2012.

\bibitem{Chatrchyan:2012ufa}
Serguei Chatrchyan et~al.
\newblock {Observation of a new boson at a mass of 125 GeV with the CMS
  experiment at the LHC}.
\newblock {\em Phys.Lett.}, B716:30--61, 2012.

\bibitem{Han:1991ia}
Tao Han and S.~Willenbrock.
\newblock {QCD correction to the $p p \to W H$ and $Z H$ total cross-sections}.
\newblock {\em Phys.Lett.}, B273:167--172, 1991.

\bibitem{Baer:1992vx}
H.~Baer, B.~Bailey, and J.F. Owens.
\newblock {O (alpha-s) Monte Carlo approach to $W +$ Higgs associated
  production at hadron supercolliders}.
\newblock {\em Phys.Rev.}, D47:2730--2734, 1993.

\bibitem{Ohnemus:1992bd}
J.~Ohnemus and W.~James Stirling.
\newblock {Order alpha-s corrections to the differential cross-section for the
  W H intermediate mass Higgs signal}.
\newblock {\em Phys.Rev.}, D47:2722--2729, 1993.

\bibitem{Kniehl:1990iva}
Bernd~A. Kniehl.
\newblock {Associated Production of Higgs and $Z$ Bosons From Gluon Fusion in
  Hadron Collisions}.
\newblock {\em Phys.Rev.}, D42:2253--2258, 1990.

\bibitem{Ciccolini:2003jy}
M.L. Ciccolini, S.~Dittmaier, and M.~Kramer.
\newblock {Electroweak radiative corrections to associated WH and ZH production
  at hadron colliders}.
\newblock {\em Phys.Rev.}, D68:073003, 2003.

\bibitem{Hamberg:1990np}
R.~Hamberg, W.L. van Neerven, and T.~Matsuura.
\newblock {A Complete calculation of the order $\alpha-s^{2}$ correction to the
  Drell-Yan $K$ factor}.
\newblock {\em Nucl.Phys.}, B359:343--405, 1991.

\bibitem{Harlander:2002wh}
Robert~V. Harlander and William~B. Kilgore.
\newblock {Next-to-next-to-leading order Higgs production at hadron colliders}.
\newblock {\em Phys.Rev.Lett.}, 88:201801, 2002.

\bibitem{Brein:2003wg}
Oliver Brein, Abdelhak Djouadi, and Robert Harlander.
\newblock {NNLO QCD corrections to the Higgs-strahlung processes at hadron
  colliders}.
\newblock {\em Phys.Lett.}, B579:149--156, 2004.

\bibitem{Brein:2012ne}
Oliver Brein, Robert~V. Harlander, and Tom~J.E. Zirke.
\newblock {vh@nnlo - Higgs Strahlung at hadron colliders}.
\newblock {\em Comput.Phys.Commun.}, 184:998--1003, 2013.

\bibitem{Ferrera:2011bk}
Giancarlo Ferrera, Massimiliano Grazzini, and Francesco Tramontano.
\newblock {Associated $WH$ production at hadron colliders: a fully exclusive
  QCD calculation at NNLO}.
\newblock {\em Phys.Rev.Lett.}, 107:152003, 2011.

\bibitem{Catani:2007vq}
Stefano Catani and Massimiliano Grazzini.
\newblock {An NNLO subtraction formalism in hadron collisions and its
  application to Higgs boson production at the LHC}.
\newblock {\em Phys.Rev.Lett.}, 98:222002, 2007.

\bibitem{Banfi:2012jh}
Andrea Banfi and Julian Cancino.
\newblock {Implications of QCD radiative corrections on high-pT Higgs
  searches}.
\newblock {\em Phys.Lett.}, B718:499--506, 2012.

\bibitem{Dawson:2012gs}
S.~Dawson, T.~Han, W.K. Lai, A.K. Leibovich, and I.~Lewis.
\newblock {Resummation Effects in Vector-Boson and Higgs Associated
  Production}.
\newblock {\em Phys.Rev.}, D86:074007, 2012.

\bibitem{Anastasiou:2008ik}
Charalampos Anastasiou, Gunther Dissertori, Fabian Stockli, and Bryan~R.
  Webber.
\newblock {QCD radiation effects on the $H \to WW \to l \nu l \nu$ signal at
  the LHC}.
\newblock {\em JHEP}, 0803:017, 2008.

\bibitem{Anastasiou:2009bt}
Charalampos Anastasiou, Guenther Dissertori, Massimiliano Grazzini, Fabian
  Stockli, and Bryan~R. Webber.
\newblock {Perturbative QCD effects and the search for a $H \to WW \to l \nu l
  \nu$ signal at the Tevatron}.
\newblock {\em JHEP}, 0908:099, 2009.

\bibitem{Stewart:2009yx}
Iain~W. Stewart, Frank~J. Tackmann, and Wouter~J. Waalewijn.
\newblock {Factorization at the LHC: From PDFs to Initial State Jets}.
\newblock {\em Phys.Rev.}, D81:094035, 2010.

\bibitem{Stewart:2010tn}
Iain~W. Stewart, Frank~J. Tackmann, and Wouter~J. Waalewijn.
\newblock {N-Jettiness: An Inclusive Event Shape to Veto Jets}.
\newblock {\em Phys.Rev.Lett.}, 105:092002, 2010.

\bibitem{Berger:2010xi}
Carola~F. Berger, Claudio Marcantonini, Iain~W. Stewart, Frank~J. Tackmann, and
  Wouter~J. Waalewijn.
\newblock {Higgs Production with a Central Jet Veto at NNLL+NNLO}.
\newblock {\em JHEP}, 1104:092, 2011.

\bibitem{Stewart:2010pd}
Iain~W. Stewart, Frank~J. Tackmann, and Wouter~J. Waalewijn.
\newblock {The Beam Thrust Cross Section for Drell-Yan at NNLL Order}.
\newblock {\em Phys.Rev.Lett.}, 106:032001, 2011.

\bibitem{Papaefstathiou:2010bw}
Andreas Papaefstathiou, Jennifer~M. Smillie, and Bryan~R. Webber.
\newblock {Resummation of transverse energy in vector boson and Higgs boson
  production at hadron colliders}.
\newblock {\em JHEP}, 1004:084, 2010.

\bibitem{Tackmann:2012bt}
Frank~J. Tackmann, Jonathan~R. Walsh, and Saba Zuberi.
\newblock {Resummation Properties of Jet Vetoes at the LHC}.
\newblock {\em Phys.Rev.}, D86:053011, 2012.

\bibitem{Banfi:2004yd}
Andrea Banfi, Gavin~P. Salam, and Giulia Zanderighi.
\newblock {Principles of general final-state resummation and automated
  implementation}.
\newblock {\em JHEP}, 0503:073, 2005.

\bibitem{Banfi:2012yh}
Andrea Banfi, Gavin~P. Salam, and Giulia Zanderighi.
\newblock {NLL+NNLO predictions for jet-veto efficiencies in Higgs-boson and
  Drell-Yan production}.
\newblock {\em JHEP}, 1206:159, 2012.

\bibitem{Bauer:2000yr}
Christian~W. Bauer, Sean Fleming, Dan Pirjol, and Iain~W. Stewart.
\newblock {An Effective field theory for collinear and soft gluons: Heavy to
  light decays}.
\newblock {\em Phys.Rev.}, D63:114020, 2001.

\bibitem{Bauer:2001yt}
Christian~W. Bauer, Dan Pirjol, and Iain~W. Stewart.
\newblock {Soft collinear factorization in effective field theory}.
\newblock {\em Phys.Rev.}, D65:054022, 2002.

\bibitem{Beneke:2002ph}
M.~Beneke, A.P. Chapovsky, M.~Diehl, and T.~Feldmann.
\newblock {Soft collinear effective theory and heavy to light currents beyond
  leading power}.
\newblock {\em Nucl.Phys.}, B643:431--476, 2002.

\bibitem{Becher:2010tm}
Thomas Becher and Matthias Neubert.
\newblock {Drell-Yan production at small $q_T$, transverse parton distributions
  and the collinear anomaly}.
\newblock {\em Eur.Phys.J.}, C71:1665, 2011.

\bibitem{Becher:2012qa}
Thomas Becher and Matthias Neubert.
\newblock {Factorization and NNLL Resummation for Higgs Production with a Jet
  Veto}.
\newblock {\em JHEP}, 1207:108, 2012.

\bibitem{Banfi:2012jm}
Andrea Banfi, Pier~Francesco Monni, Gavin~P. Salam, and Giulia Zanderighi.
\newblock {Higgs and Z-boson production with a jet veto}.
\newblock {\em Phys.Rev.Lett.}, 109:202001, 2012.

\bibitem{Bozzi:2003jy}
G.~Bozzi, S.~Catani, D.~de~Florian, and M.~Grazzini.
\newblock {The $q_T$ spectrum of the Higgs boson at the LHC in QCD perturbation
  theory}.
\newblock {\em Phys.Lett.}, B564:65--72, 2003.

\bibitem{Bozzi:2005wk}
Giuseppe Bozzi, Stefano Catani, Daniel de~Florian, and Massimiliano Grazzini.
\newblock {Transverse-momentum resummation and the spectrum of the Higgs boson
  at the LHC}.
\newblock {\em Nucl.Phys.}, B737:73--120, 2006.

\bibitem{Bozzi:2010xn}
Giuseppe Bozzi, Stefano Catani, Giancarlo Ferrera, Daniel de~Florian, and
  Massimiliano Grazzini.
\newblock {Production of Drell-Yan lepton pairs in hadron collisions:
  Transverse-momentum resummation at next-to-next-to-leading logarithmic
  accuracy}.
\newblock {\em Phys.Lett.}, B696:207--213, 2011.

\bibitem{Becher:2013xia}
Thomas Becher, Matthias Neubert, and Lorena Rothen.
\newblock {Factorization and $N^{3}LL_{p}$+NNLO predictions for the Higgs cross
  section with a jet veto}.
\newblock {\em JHEP}, 1310:125, 2013.

\bibitem{Stewart:2013faa}
Iain~W. Stewart, Frank~J. Tackmann, Jonathan~R. Walsh, and Saba Zuberi.
\newblock {Jet $p_T$ Resummation in Higgs Production at NNLL'+NNLO}.
\newblock 2013.

\bibitem{Chiu:2011qc}
Jui-yu Chiu, Ambar Jain, Duff Neill, and Ira~Z. Rothstein.
\newblock {The Rapidity Renormalization Group}.
\newblock {\em Phys.Rev.Lett.}, 108:151601, 2012.

\bibitem{Chiu:2012ir}
Jui-Yu Chiu, Ambar Jain, Duff Neill, and Ira~Z. Rothstein.
\newblock {A Formalism for the Systematic Treatment of Rapidity Logarithms in
  Quantum Field Theory}.
\newblock {\em JHEP}, 1205:084, 2012.

\bibitem{Salam:2009jx}
Gavin~P. Salam.
\newblock {Towards Jetography}.
\newblock {\em Eur.Phys.J.}, C67:637--686, 2010.

\bibitem{Catani:1993hr}
S.~Catani, Yuri~L. Dokshitzer, M.H. Seymour, and B.R. Webber.
\newblock {Longitudinally invariant $K_t$ clustering algorithms for hadron
  hadron collisions}.
\newblock {\em Nucl.Phys.}, B406:187--224, 1993.

\bibitem{Ellis:1993tq}
Stephen~D. Ellis and Davison~E. Soper.
\newblock {Successive combination jet algorithm for hadron collisions}.
\newblock {\em Phys.Rev.}, D48:3160--3166, 1993.

\bibitem{Dokshitzer:1997in}
Yuri~L. Dokshitzer, G.D. Leder, S.~Moretti, and B.R. Webber.
\newblock {Better jet clustering algorithms}.
\newblock {\em JHEP}, 9708:001, 1997.

\bibitem{Wobisch:1998wt}
M.~Wobisch and T.~Wengler.
\newblock {Hadronization corrections to jet cross-sections in deep inelastic
  scattering}.
\newblock 1998.

\bibitem{Cacciari:2008gp}
Matteo Cacciari, Gavin~P. Salam, and Gregory Soyez.
\newblock {The Anti-k(t) jet clustering algorithm}.
\newblock {\em JHEP}, 0804:063, 2008.

\bibitem{Becher:2011dz}
Thomas Becher and Guido Bell.
\newblock {Analytic Regularization in Soft-Collinear Effective Theory}.
\newblock {\em Phys.Lett.}, B713:41--46, 2012.

\bibitem{Echevarria:2012qe}
Miguel~G. Echevarria, Ahmad Idilbi, and Ignazio Scimemi.
\newblock {Definition and Evolution of Transverse Momentum Distributions}.
\newblock {\em Int.J.Mod.Phys.Conf.Ser.}, 20:92--108, 2012.

\bibitem{Becher:2007ty}
Thomas Becher, Matthias Neubert, and Gang Xu.
\newblock {Dynamical Threshold Enhancement and Resummation in Drell-Yan
  Production}.
\newblock {\em JHEP}, 0807:030, 2008.

\bibitem{Ahrens:2012qz}
Valentin Ahrens, Matthias Neubert, and Leonardo Vernazza.
\newblock {Structure of Infrared Singularities of Gauge-Theory Amplitudes at
  Three and Four Loops}.
\newblock {\em JHEP}, 1209:138, 2012.

\bibitem{Frixione:1993yp}
Stefano Frixione.
\newblock {A Next-to-leading order calculation of the cross-section for the
  production of $W^+ W^-$ pairs in hadronic collisions}.
\newblock {\em Nucl.Phys.}, B410:280--324, 1993.

\bibitem{Dawson:2013lya}
S.~Dawson, Ian~M. Lewis, and Mao Zeng.
\newblock {Threshold Resummed and Approximate NNLO results for $W^+ W^-$ Pair
  Production at the LHC}.
\newblock {\em Phys.Rev.}, D88:054028, 2013.

\bibitem{Wang:2013qua}
Yan Wang, Chong~Sheng Li, Ze~Long Liu, Ding~Yu Shao, and Hai~Tao Li.
\newblock {Transverse-Momentum Resummation for Gauge Boson Pair Production at
  the Hadron Collider}.
\newblock {\em Phys.Rev.}, D88:114017, 2013.

\bibitem{Beringer:1900zz}
J.~Beringer et~al.
\newblock {Review of Particle Physics (RPP)}.
\newblock {\em Phys.Rev.}, D86:010001, 2012.

\bibitem{Ahrens:2008qu}
Valentin Ahrens, Thomas Becher, Matthias Neubert, and Li~Lin Yang.
\newblock {Origin of the Large Perturbative Corrections to Higgs Production at
  Hadron Colliders}.
\newblock {\em Phys.Rev.}, D79:033013, 2009.

\bibitem{Campbell:2010ff}
John~M. Campbell and R.K. Ellis.
\newblock {MCFM for the Tevatron and the LHC}.
\newblock {\em Nucl.Phys.Proc.Suppl.}, 205-206:10--15, 2010.

\bibitem{ATLAS-CONF-2013-075}
{Search for associated production of the Higgs boson in the $WH \to WWW(\ast)
  \to l \nu l\nu l \nu$ and $ZH \to ZWW(\ast) \to lll \nu l \nu$ channels with
  the ATLAS detector at the LHC}.
\newblock Technical Report ATLAS-CONF-2013-075, CERN, Geneva, Jul 2013.

\bibitem{ATLAS-CONF-2013-079}
{Search for the bb decay of the Standard Model Higgs boson in associated $W/ZH$
  production with the ATLAS detector}.
\newblock Technical Report ATLAS-CONF-2013-079, CERN, Geneva, Jul 2013.

\bibitem{CMS-PAS-HIG-13-017}
{$VH$ with $H \rightarrow WW \rightarrow l \nu l\nu$ and $ V \rightarrow jj$}.
\newblock Technical Report CMS-PAS-HIG-13-017, CERN, Geneva, 2013.

\bibitem{Chatrchyan:2013zna}
Serguei Chatrchyan et~al.
\newblock {Search for the standard model Higgs boson produced in association
  with a $W$ or a $Z$ boson and decaying to bottom quarks}.
\newblock 2013.

\bibitem{Martin:2009bu}
A.D. Martin, W.J. Stirling, R.S. Thorne, and G.~Watt.
\newblock {Uncertainties on $\alpha_s$ in global PDF analyses and implications
  for predicted hadronic cross sections}.
\newblock {\em Eur.Phys.J.}, C64:653--680, 2009.

\bibitem{Botje:2011sn}
Michiel Botje, Jon Butterworth, Amanda Cooper-Sarkar, Albert de~Roeck, Joel
  Feltesse, et~al.
\newblock {The PDF4LHC Working Group Interim Recommendations}.
\newblock 2011.

\end{thebibliography}

\end{document}